\documentclass[graybox,vecphys]{svmult}

\usepackage{helvet}         
\usepackage{courier}        
\usepackage{makeidx}         
\usepackage{graphicx}        
\usepackage{multicol}        
\usepackage[bottom]{footmisc}
\ifcsname SQRT \endcsname
\else
 
\fi

\usepackage{amssymb,amsfonts,bbm}
\usepackage{cite}

\newcommand{\ensm}[1]{\ensuremath{#1}}

\renewcommand{\alph}{\ensm{\alpha}}

\newcommand{\bra}[1]{\ensm{\langle #1|}}

\newcommand{\ket}[1]{\ensm{| #1 \rangle}}

\newcommand{\be}{\begin{equation}}
\newcommand{\ee}{\end{equation}}
\newcommand{\bea}{\begin{eqnarray}}
\newcommand{\eea}{\end{eqnarray}}
\newcommand{\proj}[1]{\ket{#1}\!\bra{#1}}
\newcommand{\ot}[0]{\otimes}
\newcommand{\Tr}[1]{\mathrm{Tr}#1}
\newcommand{\tr}[1]{\mathrm{tr}#1}

\makeindex             

\begin{document}

\title*{Many-body physics from a quantum
information perspective}
\author{R. Augusiak, F. M. Cucchietti, and M. Lewenstein} 
\authorrunning{R. Augusiak, F. M. Cucchietti, and M. Lewenstein}
%
%
%
%
%

\institute{Remigiusz Augusiak \and Fernando M. Cucchietti 
\at ICFO--Institut de
Ci\`encies Fot\`oniques, Mediterranean Technology Park, 08860
Castelldefels (Barcelona), Spain, 
\and Maciej Lewenstein \at
ICFO--Institut de Ci\`encies Fot\`oniques, Mediterranean
Technology Park, 08860 Castelldefels (Barcelona), Spain,
\at ICREA--Instituci\'o Catalana de Recerca
i Estudis Avan\c{c}ats, Lluis Companys 23, 08010 Barcelona, Spain\\
\email{maciej.lewenstein@icfo.es}}



%
%
\maketitle

\abstract{
The quantum information approach to many-body physics has been very successful
in giving new insights and novel numerical methods.
In these lecture notes we take a vertical view of the subject, starting from general concepts
and at each step delving into applications or consequences of a particular topic.
We first review some general quantum information concepts like entanglement and entanglement
measures, which leads us to entanglement area laws. We then continue with one of the most famous examples of area-law
abiding states: matrix product states, and tensor product states in general.
Of these, we choose one example (classical superposition states) to introduce recent developments
on a novel quantum many-body approach: quantum kinetic Ising models.
We conclude with a brief outlook of the field.
}

\section{Introduction}
\label{sec_10_1}

There has been an explosion of interest in the interface between
quantum information (QI) and many-body systems, in particular in
the fields of condensed matter and ultracold atomic gases.
Remarkable examples are Ref.\ \cite{Jaksch99}, which proposed using
ultracold atomic gases in optical lattices for QI (and stimulated
interest in {\it distributed quantum information processing}), and
Refs. \cite{Osborne02a,Osborne02b,Osterloh02}, who discussed the
first connections between entanglement and quantum phase
transitions (QPT). Overall, the confluence of ideas has opened
fundamentally deep questions about QPT's, as well as practical
questions about how to use QI ideas in numerical simulations of
many-body quantum systems. Here, we will (partially) review these
two major themes. We will first introduce some basic
notions and tools of quantum information theory, focusing on entanglement and
entanglement measures. We shall then discuss area laws, i.e. laws that
characterize correlations and entanglement in physically relevant
many-body states, and allow to make general statements about
computational complexity of the corresponding Hamiltonians.
Afterwards, we will explore the concept of matrix product states
(MPS) and their generalizations (projected entangled pairs states,
PEPS, and tensor networks states). These states provide not only a
very useful ansatz for numerical  applications, but also a
powerful tool to understand the role of entanglement in the
quantum many-body theory. We will review one particular example of a state with
a straightforward MPS representation: the classical
superposition state. The introduction of its parent Hamiltonian will lead us to the final subject of
these lectures: quantum kinetic Ising models --- an analytically solvable
generalization of the popular classical many-body model
described by a master equation.

\section{Aspects of Quantum Information}
\label{sec_10_2}

Quantum theory contains elements that are radically different from
our everyday (``classical'') description of Nature: a most
important example are the quantum correlations present in quantum
formalism. Classically, complete knowledge of a system implies
that the sum of the information of its subsystems makes up the
total information for the whole system. In the quantum world,
this is no longer true: there exist states of composite systems
about which we have complete information, but we know nothing about
its subsystems. We may even reach paradoxical conclusions if we
apply a classical description to such ``entangled" states---whose
concept can be traced back to 1932 in manuscripts of E.
Schr{\"o}dinger.

What we have just realized during the last two decades is that
these fundamentally nonclassical  states (from hereon ``entangled
states'') can provide us with more than just paradoxes: They may
be \emph{used} to perform tasks that cannot be achieved with
classical states. As landmarks of this transformation in our view
of such nonclassical states, we mention  the spectacular
discoveries of (entanglement-based) quantum cryptography
\cite{Ekert91}, quantum dense coding \cite{Bennett92}, and quantum
teleportation \cite{Bennett93}. Even though our knowledge of
entanglement is still far from complete, significant progress has
been made in the recent years and very active research is
currently underway (for a recent and very complete review see
\cite{Horodecki09}).

In the next section, we will focus on bipartite composite systems.
We will define formally what entangled states are, present some
important criteria to discriminate entangled states from separable
ones, and show how they can be classified according to their
capability to perform some precisely defined tasks. However,
before going into details, let us introduce the notation. In what
follows we will be mostly concerned with bipartite scenarios, in
which traditionally the main roles are played by two parties
called Alice and Bob. Let  \(\mathcal{H}_A\) denote the Hilbert
space of Alice's physical system, and \(\mathcal{H}_B\) that of
Bob's. Our considerations will be restricted to finite-dimensional
Hilbert spaces, so we can set $\mathcal{H}_{A}=\mathbb{C}^{d_{A}}$
and $\mathcal{H}_{B}=\mathbb{C}^{d_{B}}$. Thus, the joint physical
system of Alice and Bob is described by the tensor product Hilbert
space \(\mathcal{H}_{AB}=\mathcal{H}_A \otimes
\mathcal{H}_{B}=\mathbb{C}^{d_{A}}\ot\mathbb{C}^{d_{B}}\).
Finally, $\mathcal{B}(\mathcal{H})$ will denote the set of bounded
linear operators from the Hilbert space $\mathcal{H}$ to
$\mathcal{H}$.

\subsection{Bipartite pure states: Schmidt decomposition}
\label{subsec:2}

We start our study with pure states, for which the concepts are simpler.
Pure states are either separable or entangled states according to the following definition:
\begin{definition}\label{SepBipPureStates}
\index{Bipartite pure states}
Consider a pure state $|\psi_{AB}\rangle$ from $\mathcal{H}_A
\otimes \mathcal{H}_{B}$. It is called separable if there exist
pure states $|\psi_{A}\rangle\in\mathcal{H}_{A}$ and
$|\psi_{B}\rangle\in\mathcal{H}_{B}$ such that
$|\psi_{AB}\rangle=|\psi_{A}\rangle\otimes |\psi_{B}\rangle$.
Otherwise we say that $\ket{\psi_{AB}}$ is entangled.
\end{definition}
The most famous examples of entangled states in $\mathcal{H}_{AB}$
are the \emph{maximally entangled states}, given by
\begin{equation}\label{MaximallyEntState}
\ket{\psi_{+}^{(d)}}=\frac{1}{\sqrt{d}}\sum_{i=0}^{d-1}\ket{i}_{A}\ot \ket{i}_{B}\qquad (d=\min\{d_{A},d_{B}\}),
\end{equation}
where the vectors $\{\ket{i}_{A}\}$ and $\{\ket{i}_{B}\}$ form
bases (in particular they can be the standard ones) in
$\mathcal{H}_{A}$ and $\mathcal{H}_{B}$, respectively. In what
follows, we also use $P_{+}^{(d)}$ to denote the
projector onto $\ket{\psi_{+}^{(d)}}$. The reason why this state
is called maximally entangled will become clear when we introduce
entanglement measures.

In pure states, the  {\it separability problem} ---
the task of judging if a given quantum state is separable --- is easy
to handle using the concept of
Schmidt decomposition:

\begin{theorem}\label{SchmidtDec}
\index{Schmidt decomposition}
Let \(\ket{\psi_{AB}}\in
\mathcal{H}_{AB}=\mathbb{C}^{d_{A}}\ot\mathbb{C}^{d_{B}}\) with
$d_{A}\leq d_{B}$. Then $\ket{\psi_{AB}}$ can be written as a
Schmidt decomposition
\begin{equation}\label{Schmidt0}
|\psi_{AB}\rangle = \sum _{i=1}^{r} \lambda_i |e_i\rangle \otimes
|f_i\rangle,
\end{equation}
where $|e_i\rangle$ and $|f_i\rangle$ form a part of an
orthonormal basis in $\mathcal{H}_{A}$ and $\mathcal{H}_{B}$,
respectively, $\lambda_i > 0$, $\sum_{i=1}^{r} \lambda_i^2=1$, and
$r\leq d_{A}$.
\end{theorem}

\begin{proof}
A generic pure bipartite state $\ket{\psi_{AB}}$ can be written in
the standard basis of $\mathcal{H}_{A}\otimes \mathcal{H}_{B}$ as
$\ket{\psi_{AB}}=\sum_{i=0} \sum_{j=0}
\alpha_{ij}\ket{i}\ot\ket{j}$, where, in general, the coefficients
$\alpha_{ij}$ form a $d_{A}\times d_{B}$ matrix $\Lambda$ obeying
$\mathrm{tr}(\Lambda^{\dagger}\Lambda)=1$. Using singular-value
decomposition, we can write $\Lambda=V D_{\Lambda}W^{\dagger}$,
where $V$ and $W$ are unitary
($V^{\dagger}V=W^{\dagger}W=\mathbbm{1}_{A}$) and $D_{\Lambda}$ is
diagonal matrix consisting of the eigenvalues $\lambda_{i}$ of
$|\Lambda|=\sqrt{\Lambda^{\dagger}\Lambda}$. Using this we rewrite
$\ket{\psi_{AB}}$ as
\begin{equation}
\ket{\psi_{AB}}=\sum_{i=0}^{d_{A}-1}\sum_{j=0}^{d_{B}-1}
\sum_{k=1}^{r}V_{ik}\lambda_{k}U_{jk}^{*}\ket{i}\ket{j},
\end{equation}
where $r\leq d_{A}\leq d_{B}$ denotes the rank of $\Lambda$. By
reshuffling terms, and defining
\(|e_{k}\rangle=\sum_{i=0}^{d_{A}-1}V_{ik}\ket{i}\) and
\(|f_{k}\rangle=\sum_{j=0}^{d_{B}-1}U^{*}_{jk}\ket{j}\)
we get the desired form [Eq.\ (\ref{Schmidt0})]. To complete
the proof, we notice that  due to the unitarity of
$V$ and $W$, vectors $\ket{e_{i}}$ and $\ket{f_{i}}$ satisfy $\langle
e_{i}|e_{j}\rangle=\langle f_{i}|f_{j}\rangle=\delta_{ij}$, and
constitute bases of $\mathcal{H}_{A}$ and
$\mathcal{H}_{B}$, respectively. In fact, $\{\lambda_{i}^{2},\ket{e_{i}}\}$ and
$\{\lambda_{i}^{2},\ket{f_{i}}\}$ are eigensystems of the first
and second subsystem of $\ket{\psi_{AB}}$. Moreover, since
$\mathrm{tr}(\Lambda^{\dagger}\Lambda)=1$ it holds that
$\sum_{i}\lambda^{2}_{i}=1$. $\blacksquare$
\end{proof}

The numbers $\lambda_{i}>0$ $(i=1,\ldots,r)$ are called \emph{the
Schmidt coefficients}, and $r$ \emph{the Schmidt rank} of
$\ket{\psi_{AB}}$. One can also notice that
$\{\lambda_{i}^{2},\ket{e_{i}}\}$ and $\{\lambda_{i}^{2},\ket{f_{i}}\}$ are
eigensystems of the first and
second subsystem of $\ket{\psi_{AB}}$, and that the Schmidt rank $r$
denotes the rank of both subsystems. Then, comparison with
definition \ref{SchmidtDec} shows that bipartite separable states
are those with Schmidt rank one. Thus, to check if a given pure
state is separable, it suffices to check the rank $r$ of one of
its subsystems. If $r=1$ (the corresponding subsystem is in a pure
state) then $\ket{\psi_{AB}}$ is separable; otherwise it is
entangled. Notice that the maximally entangled state
(\ref{MaximallyEntState}) is already written in the form
(\ref{Schmidt0}), with $r=d$ and all the Schmidt coefficients equal to
$1/\sqrt{d}$.

\subsection{Bipartite mixed states: Separable and entangled states}

The easy-to-handle separability problem in pure states complicates
considerably in the case of mixed states.  In order
to understand the distinction between separable and entangled
mixed states --- first formalized by Werner in 1989
\cite{Werner89} --- let us consider the following state
preparation procedure. Suppose that Alice and Bob are in distant
locations and can produce and manipulate any physical system in
their laboratories. Moreover, they can communicate using a
classical channel (for instance a phone line). However, they do
not have access to quantum communication channels, i.e. they are not allowed to
exchange quantum states. These two capabilities, i.e. local
operations (LO) and classical communication (CC), are frequently
referred to as LOCC.

Suppose now that in each round of the preparation scheme, Alice
generates with probability $p_{i}$ a random integer $i$
$(i=1,\ldots,K)$, which she sends to Bob.  Depending on this number,
in each round Alice prepares a pure state $\ket{e_{i}}$,
and Bob a state $\ket{f_{i}}$. After many rounds,
the result of this preparation scheme is of the form
\begin{eqnarray}
\label{eqn:sepmixed} \varrho_{AB} = \sum_{i=1}^K  p_i
|e_i\rangle\! \langle e_i| \otimes |f_i\rangle\! \langle f_i|,
\end{eqnarray}
which is the most general one that can be prepared by Alice and
Bob by means of LOCC. In this way we arrive at the formal
definition of separability in the general case of mixed states.

\begin{definition}\label{SepMixed}
\index{Bipartite mixed states}
We say that a mixed state $\varrho_{AB}$ acting on
$\mathcal{H}_{AB}$ is separable if and only
if it can be represented as a convex combination of the product of
projectors on local states as in Eq.\ (\ref{eqn:sepmixed}).
Otherwise, the mixed state is said to be entangled.
\end{definition}
The number of pure separable states $K$ necessary to decompose any
separable state according to Eq.\ (\ref{eqn:sepmixed}) is limited
by the Caratheodory theorem as \(K \leq (d_{A}d_{B} )^2\) (see Refs.
\cite{Horodecki97,Horodecki09}). No better bound is known in general, however,
for two-qubit and qubit-qutrit systems it was shown that $K\leq 4$ \cite{a} and
$K\leq 6$ \cite{b}, respectively.

By definition, entangled states cannot be prepared
locally by two parties even after communicating over a classical
channel. To prepare entangled states, the physical systems must be
brought together to interact\footnote{Due to entanglement swapping \cite{Zukowski93}, one must
suitably enlarge the notion of preparation of entangled states.
So, an entangled state between two particles can be prepared if
and only if either the two particles (call them A and B)
themselves come together to interact at a time in the past, or two
\emph{other} particles (call them C and D) do the same, with C
having interacted beforehand with A and D with B.}. Mathematically, a
nonlocal unitary operator\footnote{A unitary operator on
\(\mathcal{H}_{A}\otimes \mathcal{H}_{B}\) is said to be
``nonlocal'' if it is not of the form \(U_A \otimes U_B\), where
\(U_A\) is a unitary operator acting on \(\mathcal{H}_{A}\) and \(U_B\) acts on \(\mathcal{H}_{B}\).}
must \emph{necessarily} act on the physical system described by
\(\mathcal{H}_{A}\otimes \mathcal{H}_{B}\) to produce an
entangled state from an initial separable state.

The question whether a given bipartite state is separable or not
turns out to be quite complicated. Although the general answer to
the separability problem still eludes us, there has been
significant progress in recent years, and we will review some such
directions in the following paragraphs.

\subsection{Entanglement criteria}
\index{Entanglement criteria}

An operational necessary and sufficient criterion for detecting
entanglement still does not exist. However, over the years the
whole variety of criteria allowing for detection of entanglement
has been worked out. Below we review some of the most important
ones, while for others the reader is referred to Ref.
\cite{GuhneTothReview}. Note that, even if we do not have
necessary and sufficient separability criteria, there are
numerical checks of separability: semidefinite programming was
used to show that separability can be tested in a finite number of
steps, although this number can become too large for big systems
\cite{Doherty02,Hulpke05}. In general ---without a restriction on
dimensions--- the separability problem belongs to the NP-hard
class of computational complexity \cite{Gurvits03}.

\subsection{Partial Transposition}\index{Partial transposition}

Let us start with an easy--to--apply necessary criterion based on
the transposition map recognized by Choi \cite{Choi82} and then
independently formulated directly in the separability context by
Peres \cite{Peres96}.

Let \(\varrho_{AB}\) be a state on the product Hilbert space
\({\cal H}_{AB}\), and
$T:\mathcal{B}(\mathbb{C}^{d})\to\mathcal{B}(\mathbb{C}^{d})$ a
transposition map with respect to some real basis $\{\ket{i}\}$ in
$\mathbb{C}^{d}$, defined through $T(X)\equiv
X^{T}=\sum_{i,j}x_{ij}\ket{j}\!\bra{i}$ for any
$X=\sum_{i,j}x_{ij}\ket{i}\!\bra{j}$ from
$\mathcal{B}(\mathbb{C}^{d})$. Let us now consider an extended map
$T\ot I_{B} $ called hereafter {\it partial transposition}, where
$I_{B}$ is the identity map acting on the second subsystem. When
applied to $\varrho_{AB}$, the map $T\ot I_{B} $ transposes the
first subsystem leaving the second one untouched. More formally,
writing $\varrho_{AB}$ as
\begin{equation}
\varrho_{AB} = \sum_{i,j=0}^{d_{A}-1} \sum_{\mu,\nu=0}^{d_{B}-1}
\varrho_{ij}^{\mu\nu}|i\rangle\!\langle j|\otimes
|\mu\rangle\!\langle \nu|,
\end{equation}
where \(\{|i\rangle\}\) and \(\{|\mu\rangle\}\) are real bases
in Alice and Bob Hilbert spaces, respectively, we have

\begin{equation}
\label{eq_partial_trans} (T\ot I_{B}
)(\varrho_{AB})\equiv\varrho_{AB}^{T_A} = \sum_{i,j=0}^{d_A-1}
\sum_{\mu,\nu=0}^{d_B-1} \varrho_{ij}^{\mu\nu} |j\rangle\!\langle i|
\otimes |\mu\rangle\!\langle \nu|.
\end{equation}
Similarly, one may define partial transposition with respect to
the Bob's subsystem (denoted by $\varrho_{AB}^{T_{B}}$). Although
the partial transposition of \(\varrho_{AB}\) depends upon the
choice of the basis in which \(\varrho_{AB}\) is written, its
eigenvalues are basis independent. The applicability of the
transposition map in the separability problem can be formalized by
the following statement \cite{Peres96}.

\begin{theorem}\label{PeresTh}
If  a state \(\rho_{AB}\) is separable, then
\(\rho_{AB}^{T_{A}}\ \ge\ 0\) and \(\rho_{AB}^{T_{B}}\ \ge\ 0\).
\end{theorem}

\begin{proof}
Since \(\varrho_{AB}\) is separable, according to definition
\ref{SepMixed} it has the form (\ref{eqn:sepmixed}). Then,
performing the partial transposition with respect to the first
subsystem, we have
\begin{eqnarray}\label{Proof}
\rho_{AB}^{T_{A}} = \sum_{i=1}^{K}\ p_i\left(\proj{e_i}
\right)^{T_{A}}\otimes \proj{f_i}
= \sum_{i=1}^{K}\ p_i \proj{e_i^*} \otimes \proj{f_i}. 
\end{eqnarray}
In the second step we used that $A^{\dagger}=\left(A^*\right)^{T}$
for all $A$. The above shows that $\rho_{AB}^{T_{A}}$ is a proper
(and also separable) density matrix implying that
$\rho_{AB}^{T_{A}}\geq 0$. The same reasoning leads to the
conclusion that $\rho_{AB}^{T_{B}}\geq 0$, finishing the proof.
\(\blacksquare\)
\end{proof}
Due to the identity
$\varrho_{AB}^{T_{B}}=(\varrho_{AB}^{T_{A}})^{T}$, and the fact
that global transposition does not change eigenvalues, partial
transpositions with respect to the $A$ and $B$ subsystems are
equivalent from the point of view of the separability problem.

In conclusion, we have a simple criterion (\emph{partial
transposition criterion}) \index{Partial transposition criterion}
for detecting entanglement. More precisely, if the spectrum of one
of the partial transpositions of $\varrho_{AB}$ contains at least
one negative eigenvalue then $\varrho_{AB}$ is entangled. As an
example, let us apply the criterion to pure entangled states. If
$\ket{\psi_{AB}}$ is entangled, it can be written as
(\ref{Schmidt0}) with $r>1$. Then, the eigenvalues of
$\proj{\psi_{AB}}^{T_{A}}$ will be $\lambda_{i}^{2}$
$(i=1,\ldots,r)$ and $\pm\lambda_{i}\lambda_{j}$ $(i\neq j\;
i,j=1,\ldots,r)$. So, an entangled $\ket{\psi_{AB}}$ of Schmidt
rank $r>1$ has partial transposition with $r(r-1)/2$ negative
eigenvalues violating the criterion stated in theorem
\ref{PeresTh}.

The partial transposition criterion allows to detect in a
straightforward manner all entangled states that have
non--positive partial transposition (hereafter called \emph{NPT
states}). However, even if this is a large class of states, it
turns out that ---as pointed out in Refs.
\cite{Horodecki97,Horodecki98}--- there exist entangled states
with positive partial transposition \index{Positive partial
transposition} (called \emph{PPT states}) (cf.\ Fig.\
\ref{fig:schhil}). Moreover, the set of PPT entangled states does
not have measure zero \cite{Zyczkowski98}. It is, therefore,
important to have further independent criteria that identify
entangled PPT states. Remarkably, PPT entangled states are the
only known examples of {\it bound entangled states}, i.e., states
from which one cannot distill entanglement by means of LOCC, even
if the parties have an access to an unlimited number of copies of
the state \cite{Horodecki98,Horodecki09}. The conjecture that
there exist NPT ``bound entangled" states is one of the most
challenging open problems in quantum information theory
\cite{DiVincenzo00,Duer00}. Note also that both separable as well
as PPT states form convex sets.

Theorem \ref{PeresTh} is a necessary condition of separability in
any arbitrary dimension. However, for some special cases, the
partial transposition criterion is both a necessary and sufficient
condition for separability \cite{Horodecki96}:

\begin{theorem}\label{twoqubitsandqubitqutrit}
A state \(\varrho_{AB}\) acting on
\(\mathbb{C}^2 \otimes \mathbb{C}^2\) or \(\mathbb{C}^2 \otimes
\mathbb{C}^3\) is separable if and only if \(\varrho_{AB}^{T_{A}}\
\ge\ 0\).
\end{theorem}
We will prove this theorem later. Also, we will see that Theorem
\ref{PeresTh} is true for a whole class of maps
(of which the transposition map is only a particular example),
which also provide a {\it sufficient} criterion for separability.
Before this, let us discuss the dual characterization of
separability {\it via} entanglement witnesses.

\subsection{Entanglement Witnesses from the Hahn-Banach theorem}

Central to the concept of entanglement witnesses is the corollary
from the Hahn--Banach theorem (or Hahn--Banach
separation theorem), \index{Hahn--Banach theorem} which we will
present here limited to our needs and without proof (which the
reader can find e.g. in Ref.\ \cite{alt:1985}).
\begin{theorem}\label{HahnBanach}
Let $S$ be a convex compact set in a finite--dimensional Banach
space. Let $\rho$ be a point in this space, however, outside of the
set $S$ ($\rho \not\in S$). Then there exists a
hyperplane\footnote{A hyperplane is a linear subspace with
dimension smaller by one than the dimension of the space
itself.\index{Hyperplane}} that separates $\rho$ from $S$.
\end{theorem}
\begin{figure}[t!]
\begin{center}
\includegraphics[width=0.70\textwidth]{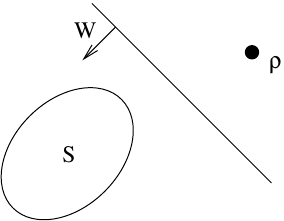}
\end{center}
\caption[Schematic picture of the Hahn-Banach theorem] {Schematic
picture of the Hahn-Banach theorem. The (unique) unit vector
orthonormal to the hyperplane can be used to define \emph{right}
and \emph{left} with respect to the hyperplane by using the sign
of the scalar product.}\label{fig:hyper}
\end{figure}
The statement of the theorem is illustrated in figure
\ref{fig:hyper}. In order to apply it to our problem let $S$
denote now the set of all separable states acting on
$\mathcal{H}_{A}\ot\mathcal{H}_{B}$. This is a convex compact
subset of the Banach space of all the linear operators
$\mathcal{B}(\mathcal{H}_{A}\ot\mathcal{H}_{B})$. The theorem
implies that for any entangled state $\varrho_{AB} $ there exists
a hyperplane separating it from $S$.

Let us introduce a coordinate system located within the hyperplane
(along with an orthogonal vector $W$ chosen so that it points
towards $S$). Then, every state $\varrho_{AB}$ can be
characterized by its ``distance" from the plane, here represented
by the Hilbert--Schmidt scalar product\footnote{Let $\mathcal{H}$
be some Hilbert space. Then the set $\mathcal{B}(\mathcal{H})$ of
linear bounded operators acting on $\mathcal{H}$ is also a Hilbert
space with the Hilbert-Schmidt scalar product $\langle
A|B\rangle=\tr(A^{\dagger} B)$ ($A,B\in\mathcal{B}(\mathcal{H})$).
}. According to our choice of the coordinate system (see Fig.\
\ref{fig:hyper}), for any such hyperplane $W$ every separable
state has a positive ``distance", while there are some entangled
states with a negative ``distance". More formally, theorem
(\ref{HahnBanach}) implies the following seminal result \cite{Horodecki96}.
\begin{theorem}\label{WitnessTh}
Let $\varrho_{AB}$ be some entangled state
acting on $\mathcal{H}_{AB}$. Then there exists a Hermitian
operator $W\in \mathcal{B}(\mathcal{H}_{A}\ot\mathcal{H}_{B})$
such that $\tr(\varrho_{AB}W)<0$ and $\tr(\sigma_{AB}W)\geq 0$ for
all separable
$\sigma_{AB}\in\mathcal{B}(\mathcal{H}_{A}\ot\mathcal{H}_{B})$.
\end{theorem}
It is then clear that all the operators $W$ representing such
separating hyperplanes deserve special attention as they are
natural candidates for entanglement detectors. That is, given some
Hermitian $W$, if $\tr(W\varrho_{AB})<0$ and simultaneously
$\tr(W\sigma_{AB})\geq 0$ for all separable $\sigma_{AB}$, we know
that $\varrho_{AB}$ is entangled. One is then tempted to introduce
the following definition \cite{TerhalPLA00}.
\begin{definition}
We call the Hermitian operator $W$ an entanglement witness
if\linebreak $\tr(W\sigma_{AB})\geq 0$ for all separable
$\sigma_{AB}$ and there exists an entangled state $\varrho_{AB}$
such that $\tr(W\varrho_{AB})<0$.
\end{definition}

\begin{example}
Let us discuss how to construct entanglement witnesses for all NPT
states. If $\varrho_{AB}$ is NPT then its partial transposition
has at least one negative eigenvalue. Let $\ket{\psi_{i}}$ denote
the eigenstates of $\varrho^{T_{B}}_{AB}$ corresponding to its
negative eigenvalues $\lambda_{i}<0$. Then the Hermitian operator
$W_{i}=\proj{\psi_{i}}^{T_{B}}$ has negative mean value on
$\varrho_{AB}$, i.e.,
$\tr(\varrho_{AB}\proj{\psi_{i}}^{T_{B}})=\tr(\varrho_{AB}^{T_{B}}\proj{\psi_{i}})=\lambda_{i}<0$.
Simultaneously, using the identity $\tr(AB^{T})=\tr(A^{T}B)$
obeyed by any pair of matrices $A$ and $B$, it is straightforward
to verify that $\tr(W_{i}\sigma_{AB})\geq 0$ for all $i$ and
separable $\sigma_{AB}$. One notices also that any convex
combination of $W_{i}$ and in particular $\varrho_{AB}^{T_{B}}$
itself are also entanglement witnesses.
\end{example}

Let us comment shortly on the properties of entanglement
witnesses. First, it is clear that they have negative eigenvalues,
as otherwise their mean value on all entangled states would be
positive. Second, since entanglement witnesses are Hermitian, they
can be treated as physical observables --- which means that
separability criteria based on entanglement witnesses are
interesting from the experimental point of view. Third, even if
conceptually easy, entanglement witnesses depend on states in the
sense that there exist entangled states that are only detected by
different witnesses. Thus, in principle, the knowledge of all
entanglement witnesses is necessary to detect all entangled
states.


\subsection{Positive maps and the entanglement problem}

Transposition is not the only map that can be used to deal with
the separability problem. It is rather clear that the statement of
theorem \ref{PeresTh} remains true if, instead of the
transposition map, one uses any map that when applied to a
positive operator gives again a positive operator (a {\it positive
map}). Remarkably, as shown in Ref.\ \cite{Horodecki96}, positive
maps give not only necessary but also sufficient conditions for
separability and entanglement detection. Moreover, {\it via} the
Jamio\l{}kowski-Choi isomorphism, theorem \ref{WitnessTh} can be
restated in terms of positive maps. To see this in more detail we
need to review a bit of terminology.

We say that a map \(\Lambda: \mathcal{B} (\mathcal{H}_{A})
\rightarrow \mathcal{B}(\mathcal{H}_{B})\) is linear if
$\Lambda(\alpha X+\beta Y)=\alpha\Lambda(X)+\beta\Lambda(Y)$ for
any pair of operators $X,Y$ acting on $\mathcal{H}_{A}$ and
complex numbers $\alpha,\beta$. We also say that $\Lambda$ is
Hermiticity--preserving (trace--preserving) if
$\Lambda(X^{\dagger})=[\Lambda(X)]^{\dagger}$
($\tr[\Lambda(X)]=\tr(X)$) for any Hermitian
$X\in\mathcal{B}(\mathcal{H}_{A})$.

\begin{definition}
A linear map
$\Lambda:\mathcal{B}(\mathcal{H}_{A})\rightarrow\mathcal{B}(\mathcal{H}_{B})$
is called positive if for all positive
\(X\in\mathcal{B}(\mathcal{H}_{A})\) the operator
\(\Lambda(X)\in\mathcal{B}(\mathcal{H}_{B})\) is positive.
\end{definition}
As every Hermitian operator can be written as a difference of two
positive operators, any positive map is also
Hermiticity--preserving. On the other hand, a positive map does
not have to be necessarily trace--preserving.

It follows immediately from the above definition that positive
maps applied to density matrices give (usually unnormalized)
density matrices. One could then expect that positive maps are
sufficient to describe all quantum operations (as for instance
measurements). This, however, is not enough, as it may happen that
the considered system is only part of a larger one and we must
require that any quantum operation on our system leaves the global
system in a valid physical state. This requirement leads us to the
notion of completely positive maps:

\begin{definition}\label{ComplPosMaps}
\index{Completely positive map} Let
\(\Lambda:\mathcal{B}(\mathcal{H}_{A})\rightarrow
\mathcal{B}(\mathcal{H}_{B})\) be a positive map and let
\(I_{d}:M_{d}(\mathbb{C})\rightarrow M_{d}(\mathbb{C})\) denote an
identity map. Then, we say that \(\Lambda\) is completely positive
if for all \(d\) the extended map \(I_{d}\otimes \Lambda\) is
positive.
\end{definition}
Let us illustrate the above definitions with some
examples.
\begin{example}
(Hamiltonian evolution of a quantum state) Let
$\mathcal{H}_{A}=\mathcal{H}_{B}=\mathcal{H}$ and let
$\Lambda_{U}:\mathcal{B}(\mathcal{H})\rightarrow
\mathcal{B}(\mathcal{H})$ be defined as
$\Lambda_{U}(X)=UXU^{\dagger}$ for any
$X\in\mathcal{B}(\mathcal{H})$, with $U$ being some unitary
operation acting on $\mathcal{H}$. Since unitary operations do not
change eigenvalues when applied to $X$, it is clear that
$\Lambda_{U}$ is positive for any such $U$. Furthermore,
$\Lambda_{U}$ is completely positive: an application of the
extended map $I_{d}\otimes \Lambda_{U}$ to
$X\in\mathcal{B}(\mathcal{H}\otimes\mathcal{H})$ gives
$(I_{d}\otimes \Lambda_{U}) (X)=(\mathbbm{1}_{d}\ot U) X
(\mathbbm{1}_{d}\ot U)^{\dagger}$, where $\mathbbm{1}_{d}$ denotes
identity acting on $\mathcal{H}$. Therefore, the extended unitary
$\widetilde{U}=\mathbbm{1}_{d}\ot U$ is also unitary. Thus, if
$X\geq 0$, then $\widetilde{U}X\widetilde{U}^{\dagger}\geq0$.
The commonly known example of $\Lambda_{U}$ is the unitary
evolution of a quantum state
$\varrho(t)=U(t)\varrho(0)[U(t)]^{\dagger}=\Lambda_{U(t)}(\varrho(0))$.
\end{example}

\begin{example}
(Transposition map) The second example of a linear map is the
already considered transposition map $T$. It is easy to check that
$T$ is Hermiticity and trace--preserving. However, the previously
discussed example of partially transposed pure entangled states
shows that it cannot be completely positive.
\end{example}

To complete the characterization of positive and completely
positive maps let us just mention the {\it
Choi--Kraus--Stinespring representation}. Recall first that any
linear Hermiticity--preserving (and so positive) map
$\Lambda:\mathcal{B}(\mathbb{C}^{d})\rightarrow
\mathcal{B}(\mathbb{C}^{d})$ can be represented as
\cite{GoriniKossakowskiSudarshan76}:
\begin{equation}\label{PosMap}
\Lambda(X)=\sum_{i=1}^{k}\eta_{i}V_{i}X V_{i}^{\dagger},
\end{equation}
where $k\leq d^{2}$, $\eta_{i}\in\mathbb{R}$, and
$V_{i}:\mathbb{C}^{d}\rightarrow \mathbb{C}^{d}$ are orthogonal in
the Hilbert--Schmidt scalar product
$\tr(V_{i}^{\dagger}V_{j})=\delta_{ij}$. In this representation,
completely positive maps are those (and only those) that have
$\eta_{i}\geq 0$ for all $i$. As a result, by replacing
$W_{i}=\sqrt{\eta_{i}}\,V_{i}$ (which preserves the orthogonality
of $W_{i}$), we arrive at the aforementioned form for completely
positive maps \cite{Choi75,Kraus83,Stinespring55}.
\begin{theorem}\label{KrausForm}
A linear map $\Lambda:\mathcal{B}(\mathbb{C}^{d})\rightarrow
\mathcal{B}(\mathbb{C}^{d})$ is completely positive iff admits the
Choi--Kraus--Stinespring form
\begin{equation}
\Lambda(X)=\sum_{i=1}^{k}V_{i}X V_{i}^{\dagger},
\end{equation}
where $k\leq d^{2}$ and $V_{i}:\mathbb{C}^{d}\rightarrow
\mathbb{C}^{d}$, called usually Kraus operators, are orthogonal in
the Hilbert--Schmidt scalar product.
\end{theorem}

Finally, let us recall the so--called Choi--Jamio{\l}kowski
isomorphism \cite{Jamiolkowski72,Choi75}: every linear operator
$X$ acting on $\mathbb{C}^{d}\ot\mathbb{C}^{D}$ can be represented
as $X=(I\ot\Lambda)(P_{+}^{(d)})$ with some linear map
$\Lambda:\mathcal{B}(\mathbb{C}^{d})
\rightarrow\mathcal{B}(\mathbb{C}^{D})$. With this isomorphism,
entanglement witnesses correspond to positive maps. Notice also
that the dual form of this isomorphism reads
$\Lambda(X)=\tr_{B}[W(\mathbbm{1}_{A}\ot X^{T})]$).

Equipped with new definitions and theorems, we can now continue
with the relationship between positive maps and the separability
problem. It should be clear by now that theorem \ref{PeresTh} is
just a special case of a more general necessary condition for
separability: if $\varrho_{AB}$ acting on
\(\mathcal{H}_{A}\ot\mathcal{H}_{B}\) is separable, then \((I\ot
\Lambda)(\varrho_{AB})\) is positive for any positive map
$\Lambda$. In a seminal paper in 1996 \cite{Horodecki96},
the Horodeckis showed that positive maps also give a
\emph{sufficient} condition for separability. More precisely, they
proved the following \cite{Horodecki96}:

\begin{theorem}
A state $\rho_{AB}\in \mathcal{B}
(\mathbb{C}^{d_{A}} \otimes \mathbb{C}^{d_{B}})$ is separable if
and only if the condition
\begin{equation}\label{CondSep}
(I\otimes\Lambda)(\rho_{AB})\geq 0.
\end{equation}
holds for all positive maps
$\Lambda:\mathcal{B}(\mathbb{C}^{d_{B}})\rightarrow
\mathcal{B}(\mathbb{C}^{d_{A}})$.
\end{theorem}
\begin{proof}
The ``only if" part goes along exactly the same lines as proof of
theorem \ref{PeresTh}, where instead of the
transposition map we put $\Lambda$.
On the other hand, the ``if" part is much more involved. Assuming
that $\varrho_{AB}$ is entangled, we show that there exists
a positive map $\Lambda:\mathcal{B}(\mathbb{C}^{d_{B}})\rightarrow
\mathcal{B}(\mathbb{C}^{d_{A}})$ such that
$(I\ot\Lambda)(\varrho_{AB})\ngeq 0$. For this we can use theorem
\ref{WitnessTh}, which says that for any entangled $\varrho_{AB}$
there always exists entanglement witness $W$ detecting it, i.e.,
$\tr(W\varrho_{AB})<0$.
Denoting by
$L:\mathcal{B}(\mathbb{C}^{d_{A}})\to\mathcal{B}(\mathbb{C}^{d_{B}})$
a positive map corresponding to the witness $W$ via the the
Choi--Jamio{\l}kowski isomorphism, i.e., $W=(I\ot
L)(P_{+}^{(d_{A})})$, we can rewrite this condition as
\begin{equation}
\tr[(I\ot L)(P_{+}^{(d_{A})})\varrho_{AB}]<0.
\end{equation}
As $L$ is positive it can be represented as in Eq.\ (\ref{PosMap}),
and hence the above may be rewritten as $\Tr[P_{+}^{(d_{A})}(I\ot
L^{\dagger})(\varrho_{AB})]$ with
$L^{\dagger}:\mathcal{B}(\mathbb{C}^{d_{B}})\rightarrow\mathcal{B}(\mathbb{C}^{d_{A}})$
called the dual map of $L$. One immediately checks that dual maps
of positive maps are positive. This actually finishes the proof
since we showed that there exists a positive map
$\Lambda=L^{\dagger}$ such that $(I\ot\Lambda)(\varrho_{AB})\ngeq
0$. $\blacksquare$
\end{proof}
In conclusion, we have two equivalent characterizations of
separability in bipartite systems, in terms of either entanglement
witnesses or positive maps. However, on the level of a
particular entanglement witness and the corresponding map, both
characterizations are no longer equivalent. This is because
usually maps are stronger in detection than entanglement witnesses
(see Ref.\ \cite{Horodecki01c}). A good example comes from the two
qubit case. On one hand, theorem \ref{twoqubitsandqubitqutrit}
tell us that the transposition map detects all the two--qubit
entangled states. On the other hand, it is clear that the
corresponding witness, the so--called swap operator (see Ref.
\cite{Werner89}) $V=P_{+}^{(2)\Gamma}$ does not detect all
entangled states --- as for instance $\tr(P_{+}^{(2)}V)\geq0$.

Let us also notice that an analogous theorem was proven in Ref.
\cite{Horodecki01c}, which gave a characterization of the set of
the fully separable multipartite states
\begin{equation}
\varrho_{A_{1}\ldots A_{N}}=\sum_{i}p_{i}\varrho_{A_{1}}^{(i)}\ot
\ldots\ot\varrho_{A_{N}}^{(i)}
\end{equation}
in terms of multipartite entanglement witnesses. Here,
however, instead of positive maps one deals with maps which are
positive on products of positive operators.

\subsection{Positive maps and entanglement witnesses: further characterization and examples}

We discuss here the relationship between
positive maps (or the equivalent entanglement witnesses)
and the separability problem.

\begin{definition}\label{DecMaps}
Let
$\Lambda:\mathcal{B}(\mathcal{H}_{A})\rightarrow\mathcal{B}(\mathcal{H}_{B})$
be a positive map. We call it decomposable if it admits the
form\footnote{By $\Lambda_1\circ\Lambda_2$ we denote the
composition of two maps $\Lambda_i$ $(i=1,2)$, i.e., a map that
acts on a given operator $X$ as
$\Lambda_1\circ\Lambda_2(X)=\Lambda_1(\Lambda_2(X))$.}
$\Lambda=\Lambda_{1}^{\mathrm{CP}}+\Lambda_{2}^{\mathrm{CP}}\circ
T$, where $\Lambda_{i}^{\mathrm{CP}}$ $(i=1,2)$ are some
completely positive maps. Otherwise $\Lambda$ is called
indecomposable.
\end{definition}
It follows from this definition that
decomposable maps are useless for detection of PPT
entangled states. To see this explicitly, assume that
$\varrho_{AB}$ is PPT entangled. Then it holds that
$(I\ot\Lambda)(\varrho_{AB})=
(I\ot\Lambda_{1}^{\mathrm{CP}})(\varrho_{AB})+
(I\ot\Lambda_{2}^{\mathrm{CP}})(\varrho_{AB}^{T_{B}})=
(I\ot\Lambda_{1}^{\mathrm{CP}})(\varrho_{AB})+
(I\ot\Lambda_{2}^{\mathrm{CP}})(\widetilde{\varrho}_{AB})$, where
$\widetilde{\varrho}_{AB}=\varrho_{AB}^{T_{B}}$ is some quantum state.
Since $\Lambda_{i}^{\mathrm{CP}}$ are completely positive, both terms
are positive and thus $(I\ot\Lambda)(\varrho_{AB})\geq 0$ for any decomposable
$\Lambda$ and PPT entangled $\varrho_{AB}$.

The simplest example of a decomposable map is the transposition
map, with both $\Lambda_{i}^{\mathrm{CP}}$ $(i=1,2)$ being just
the identity map. It is then clear that, from the point of view of
entanglement detection, the transposition map is also the most
powerful example of a decomposable map. Furthermore, as shown by
Woronowicz \cite{Woronowicz}, all positive maps from
$\mathcal{B}(\mathbb{C}^{2})$ and $\mathcal{B}(\mathbb{C}^{3})$ to
$\mathcal{B}(\mathbb{C}^{2})$ are decomposable. Therefore, the
partial transposition criterion is necessary and sufficient in
two-qubit and qubit-qutrit systems as stated in theorem
\ref{twoqubitsandqubitqutrit}.

Using the Jamio\l{}kowski-Choi isomorphism we can check
the form of entanglement witnesses corresponding to the decomposable
positive maps. One immediately sees that they
can be written as $W=P+Q^{T_{B}}$,
with $P$ and $Q$ being some positive operators. Following
the nomenclature of positive maps, such witnesses are called decomposable.

\begin{figure}[t!]
\begin{center}
\includegraphics[width=0.70\textwidth]{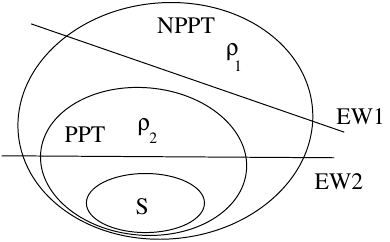}
\end{center}
\caption[Comparison of a nondecomposable entanglement witness with
a decomposable one] {Schematic view of the Hilbert-space with two
states $\rho_1$ and $\rho_2$ and two witnesses $EW1$ and $EW2$.
$EW1$ is a decomposable EW, and it detects only  NPT states like
$\rho_1$. $EW2$ is an indecomposable EW, and it detects also some
PPT states like $\rho_2$. Note that none of the witnesses detect
\emph{all} entangled states.}\label{fig:schhil}
\end{figure}

It is then clear that PPT entangled states can only be detected by
indecomposable maps, or, equivalently indecomposable
entanglement witnesses (cf.\ Fig.\ \ref{fig:schhil}). Still,
however, there is no criterion that allows to judge unambiguously
if a given PPT state is entangled.

%

To support the above discussion, we give particular
examples of positive maps and corresponding entanglement
witnesses.
\begin{example}
Let $\Lambda_{r}:\mathcal{B}(\mathbb{C}^{d})\rightarrow
\mathcal{B}(\mathbb{C}^{d})$ be the so-called {\it reduction map}
map defined through $\Lambda_{r}(X)=\tr(X)\mathbbm{1}_{d}-X$ for
any $X\in\mathcal{B}(\mathbb{C}^{d})$. It was introduced in Ref.
\cite{TanahashiTomiyama88} and considered first in the
entanglement context in Refs. \cite{Horodecki00,Cerf99}. One
immediately finds that $\Lambda_{r}$ is positive, but not
completely positive, as it detects entanglement of $P_{+}^{(d)}$.
Moreover, $\Lambda_{r}=\Lambda^{\mathrm{CP}}\circ T$, where
$\Lambda^{\mathrm{CP}}$ is a completely positive map with Kraus
operators (cf. theorem \ref{KrausForm}) given by
$V_{ij}=|i\rangle\!\langle j|-|j\rangle\!\langle i|$
$(i<j,\;i,j=0,\ldots,d-1)$, meaning that the reduction map is
decomposable.
\end{example}
\begin{example}
Let $\Lambda_{ext}^{U}:\mathcal{B}(\mathbb{C}^{d})\rightarrow
\mathcal{B}(\mathbb{C}^{d})$ be the so-called extended reduction
map \cite{BreuerPRL06,HallJPA06} defined by
$\Lambda_{ext}^{U}(X)=\tr(X)\mathbbm{1}_{d}-X-UX^{T}U^{\dagger}$,
where $U$ obeys $U^{T}=-U$ and $U^{\dagger}U\leq \mathbbm{1}_{d}$.
It is obviously positive but not completely positive. However,
unlike the reduction map, this one is indecomposable as examples
of PPT entangled states detected by $\Lambda_{ext}^{U}$ can be
found \cite{BreuerPRL06,HallJPA06}.
\end{example}

Let us summarize our considerations with the following two
theorems. First, using the definitions of decomposable and
indecomposable entanglement witnesses, we can restate the
consequences of the Hahn-Banach theorem in several ways 
\cite{Woronowicz76,Choi82,Horodecki96,Lewenstein00,Terhal01}:
\begin{theorem}
The following statements hold.
\begin{enumerate}
\item A state $\rho_{AB}$ is entangled iff there exists an entanglement witness $W$ such that $\tr(W\rho_{AB}) < 0$.
\item A state $\rho_{AB}$ is PPT entangled iff there exists an indecomposable entanglement witness $W$
such that $\tr(W\rho_{AB}) < 0$.
\item A state $\sigma_{AB}$ is separable iff  $\tr(W\sigma_{AB})\geq 0$ for all entanglement witnesses.
\end{enumerate}
\end{theorem}
Notice that the Jamio\l{}kowski-Choi isomorphism between positive
maps and entanglement witnesses allows to rewrite immediately the
above theorem in terms of positive maps. From a theoretical point
of view, the theorem is quite powerful. However, it does not give
any insight on how to construct for a given state $\rho$, the
appropriate witness operator.

Second, the relations between maps and witnesses can be collected
as follows 
\cite{Jamiolkowski72,Woronowicz76,Horodecki96,Lewenstein00,Terhal01}.
\begin{theorem}
Let $W$ be a Hermitian operator and $\Lambda_{W}$ map defined as
$\Lambda_{W}(X)=\tr_{B}[W(\mathbbm{1}_{A}\ot X^{T})]$.
Then the following statements hold.
\begin{enumerate}
\item $W \geq 0$ iff $\Lambda_{W}$ is a completely positive map.
\item $W$ is an entanglement witness iff $\Lambda_{W}$ is a positive map.
\item $W$ is a decomposable entanglement witness iff $\Lambda_{W}$ is decomposable map.
\end{enumerate}
\end{theorem}

\subsection{Entanglement measures}

The criteria discussed above allow to check if a given state
$\varrho_{AB}$ is entangled. However, in general they do not tell
us directly {\em how much} $\varrho_{AB}$ is entangled. In what
follows we discuss several methods to quantify entanglement of
bipartite states. This quantification is necessary, at least
partly because entanglement is viewed as a resource in quantum
information theory. There are several complementary ways to
quantify entanglement (see Refs.
\cite{Bennett96b,Vedral97,DiVincenzo98,Laustsen03,Nielsen99,Vidal00,Jonathan99,Horodecki04,Horodecki01b,Plenio07,Horodecki09}
and references therein). We will present here three possible ways
to do so.

Let us just say few words about the definition of entanglement
measures\footnote{For a more detailed axiomatic description, and
other properties of entanglement measures, the reader is
encouraged to consult, e.g., Refs.\
\cite{Horodecki01b,Plenio07,Horodecki09}.}. The main ingredient in
this definition is the monotonicity under LOCC operations. More
precisely, if $\Lambda$ denotes some LOCC operation, and $E$ our
candidate for the entanglement measure, $E$ has to satisfy
\begin{equation}\label{monot1}
E(\Lambda(\varrho))\leq E(\varrho)
\end{equation}
or
\begin{equation}\label{monot2}
\sum_{i}p_{i}E(\varrho_{i})\leq E(\varrho),
\end{equation}
where $\varrho_{i}$ are states resulting from the LOCC operation
$\Lambda$ appearing with probabilities $p_{i}$ (as in the case of
e.g. projective measurements). Both requirements follow from the
very intuitive condition saying that entanglement should not
increase under local operations and classical communication. It
follows also that if $E$ is convex, then the condition
(\ref{monot2}) implies (\ref{monot1}), but not vice versa ---
therefore (\ref{monot2}) gives a stronger condition for the
monotonicity. For instance, the three examples of measures
presented below satisfy this condition. Finally, notice that from
the monotonicity under LOCC operations one also concludes that $E$
is invariant under unitary operations, and gives a constant value
on separable states (see e.g. Ref.\ \cite{Horodecki09}).

\subsubsection{Entanglement of formation}

Consider a bipartite pure state
\(\left|\psi_{AB}\right\rangle\in\mathbb{C}^{d_{A}}\ot\mathbb{C}^{d_{B}}\)
shared between Alice and Bob. As shown by Bennett \emph{et al.}
\cite{Bennett96a}, given \(n E(\ket{\psi_{AB}})\) copies of the
maximally entangled state, Alice and Bob can by LOCC transform
them into \(n\) copies of \(\left|\psi_{AB}\right\rangle\), if
\(n\) is large. Here
\begin{equation}\label{eq_ent_pure}
E(\ket{\psi_{AB}})=S(\varrho_A)= S(\varrho_B)
\end{equation}
with \(\varrho_A\) and \(\varrho_B\) being the local density
matrices of \(\left|\psi_{AB}\right\rangle\) and \(S(\rho)\)
stands for the von Neumann entropy of \(\rho\) given by $S(\rho) =
- \tr\rho \log_{2} \rho$. It clearly follows from theorem
\ref{SchmidtDec} that $E$ is zero iff $\ket{\psi_{AB}}$ is
separable, while its maximal value $\log_2\min\{d_{A},d_{B}\}$ is
attained for the maximally entangled states
(\ref{MaximallyEntState}).

For the two-qubit maximally entangled state
$\ket{\psi_{+}^{(2)}}$, the function $E$ gives one: an amount
of entanglement also called {\it ebit}. With this terminology,
one can say that $\ket{\psi_{AB}}$ has
$E(\ket{\psi_{AB}})$ ebits. Since \(E(\psi_{AB})\) is the number
of singlets required to prepare a copy of the state
\(\left|\psi_{AB}\right\rangle\), it is called {\it entanglement
of formation} of \(\left|\psi_{AB}\right\rangle\). We are
therefore using the amount of entanglement of the singlet state as
our unit of entanglement.

Following Ref. \cite{Bennett96b}, let us now extend the definition
of entanglement of formation to all bipartite states. By
definition, any mixed state is a convex combination of pure
states, i.e., $\varrho=\sum_{i}p_{i}\proj{\psi_{i}}$, where
probabilities $p_{i}$ and pure states (not necessarily orthogonal)
$\ket{\psi_{i}}$ constitute what is called an ensemble. A
particular example of such an ensemble is the eigendecomposition
of $\varrho$. Thus, it could be tempting to define the
entanglement of formation of $\varrho$ as an averaged cost of
producing pure states from the ensemble, i.e.,
$\sum_{i}p_{i}E(\ket{\psi_{i}})$. One knows, however, that there
exist an infinite number of ensembles realizing any given
$\varrho$. A natural solution is then to minimize the above
function over all such ensembles --- with which we arrive at the
definition of entanglement of formation for mixed states
\cite{Bennett96b}:
\begin{equation}
E(\varrho_{AB})=\min_{\{p_{i},\ket{\psi_{i}}\}}\sum_{i}p_{i}E(\ket{\psi_{i}}),
\end{equation}
with the minimum taken over all ensembles
$\{p_{i},\ket{\psi_{i}}\}$ such that
$\sum_{i}p_{i}\proj{\psi_{i}}=\varrho_{AB}$.

In general, the above minimization makes the calculation of
entanglement of formation extremely difficult. Nevertheless, it
was determined for two-qubits \cite{Hill97,Wootters98}, or states
having some symmetries, as the so--called isotropic \cite{Terhal00}
and Werner \cite{VollbrechtWerner} states. In the first case it
amounts to
\begin{equation}
E_F(\varrho_{AB}) = H\left( \frac{1+ \sqrt{1 -
C^2(\varrho_{AB})}}{2} \right),
\end{equation}
where \(H(x) = -x\log_2 x - (1-x)\log_2 (1-x)\) is the binary
entropy function. The function $C$ is given by
\begin{equation}\label{Concurrence2Q}
C(\varrho_{AB}) = \max\left\{0,
\lambda_1 - \lambda_2 - \lambda_3 - \lambda_4 \right\}.
\end{equation}
with $\lambda_1,\ldots,\lambda_4$ the eigenvalues of the Hermitian
matrix \([(\varrho_{AB})^{1/2} \tilde{\varrho}_{AB}
(\varrho_{AB})^{1/2}]^{1/2}\) in decreasing order, and
\(\tilde{\varrho}_{AB}= \sigma_y \otimes \sigma_y \varrho_{AB}^*
\sigma_y \otimes \sigma_y\). Note that the complex conjugation
over \(\varrho\) is taken in the \(\sigma_z\) eigenbasis, and
$\sigma_{y}$ denotes the well-known Pauli matrix\footnote{In the
standard basis $\sigma_{y}$ is given by
$\sigma_{y}=-i|0\rangle\langle1|+i|1\rangle\langle 0|$.}.
The function $C$, called {\it concurrence}, can also be used to
quantify entanglement of more general quantum states. Although Eq.
(\ref{Concurrence2Q}) gives the explicit form of concurrence only for two-qubit
states, it can also be defined for arbitrary bipartite
states --- as we shall discuss in the following section.

\subsubsection{Concurrence}

For any $\ket{\psi_{AB}}\in
\mathbb{C}^{d_{A}}\ot\mathbb{C}^{d_{B}}$ we define concurrence as
$C(\ket{\psi_{AB}})=\sqrt{2(1-\tr\varrho_{r}^{2})}$ where
$\varrho_{r}$ is one of the subsystems of $\ket{\psi_{AB}}$ (note
that the value of $C$ does not depend on the choice of subsystems)
\cite{Rungta01}. In the case $d_{A}=d_{B}=d$, one sees that its
value for pure states ranges from $0$ for separable states to
$\sqrt{2(1-1/d)}$ for the maximally entangled state.

The extension to mixed states goes in exactly the same way as in the case
of entanglement of formation,
\begin{equation}
C(\varrho_{AB})=\min_{\{p_{i},\ket{\psi_{i}}\}}\sum_{i}p_{i}C(\ket{\psi_{i}}),
\end{equation}
where again the minimization is taken over all the ensembles that realize
$\varrho_{AB}$. For the same reason, as in the case of EOF,
concurrence is calculated analytically only in few instances like two-qubit
states \cite{Hill97,Wootters98} and isotropic states
\cite{RungtaCaves03}.

Seemingly, the only difference between $E$ and $C$ lies in
the function taken to define both measures for pure states.
However, the way concurrence is defined enables one to determine
it experimentally for pure states
\cite{AolitaMintertPRL06,WalbornNat06}, provided that two copies
of the state are available simultaneously.

\subsubsection{Negativity and logarithmic negativity}

Based on the previous examples of entanglement measures, one may
get the impression that all of them are difficult to determine.
Even if this is true in general, there are entanglement measures
that can be calculated for arbitrary states. The examples we
present here are {\it negativity} and {\it logarithmic
negativity}. The first one is defined as
\cite{Zyczkowski98,Vidal02b}:
\begin{equation}
N(\varrho_{AB})=\frac{1}{2}\left(\left\|\varrho_{AB}^{\Gamma}\right\|-1\right).
\end{equation}
The calculation of $N$ even for mixed states reduces to
determination of eigenvalues of $\varrho_{AB}^{T_{B}}$, and
amounts to the sum of the absolute values of negative eigenvalues
of $\varrho_{AB}^{T_{B}}$. This measure has a disadvantage:
partial transposition does not detect PPT entangled states;
therefore $N$ is zero not only for separable states but also for
all PPT states.

The logarithmic negativity is defined as \cite{Vidal02b}:
\begin{equation}\label{LogNeg}
E_{N}(\varrho_{AB})=\log_2\left\|\varrho^{\Gamma}_{AB}\right\|=\log_2[2N(\varrho_{AB})+1].
\end{equation}
It was shown in Ref.\ \cite{Plenio05a} that it satisfies condition
(\ref{monot2}). Moreover, logarithmic negativity is additive,
i.e.,
$E(\varrho_{AB}\ot\sigma_{AB})=E(\varrho_{AB})+E(\sigma_{AB})$ for
any pair of density matrices $\varrho_{AB}$ and $\sigma_{AB}$,
which is a desirable feature. However, this comes at a cost:
$E_{N}$ is not convex \cite{Plenio05a}. Furthermore, for the same
reason as negativity it cannot be used to quantity entanglement of
PPT entangled states. Finally, let us notice that these measures
range from zero for separable states, to $(d-1)/2$ for negativity
and $\log_2 d$ for logarithmic negativity.

\section{Area laws}

Area laws play a very important role in many areas of physics,
since generically relevant states of physical systems described by
local Hamiltonians (both quantum and classical) fulfill them. This
goes back to the seminal work on the free Klein--Gordon field
\cite{Bombelli86,Srednicki93}, where it was suggested that the
area law of geometric entropy might be related to the physics of
black holes, and in particular the Bekenstein-Hawking entropy that
is proportional to the area of the black hole surface
\cite{Bekenstein73,Bekenstein04,Hawking74}. The related {\it
holographic principle} \cite{Bousso02} says that information about
a region of space can be  represented by a theory which lives on a
boundary of that region. In recent years there has been a wealth
of studies of area laws, and there are excellent reviews
\cite{EisertCramerPlenioReview} and special issues
\cite{Calabrese10} about the subject. As pointed out by the
authors of Ref.\ \cite{EisertCramerPlenioReview}, the interest in
area laws is particularly motivated by the four following issues:
\begin{itemize}
\item The holographic principle and the entropy of black holes,
\item Quantum correlations in many-body systems,
\item Computational complexity of quantum many-body systems,
\item Topological entanglement entropy as an indicator of topological order in certain many-body systems
\end{itemize}

\subsection{Mean entanglement of bipartite states}

Before we turn to the area laws for physically relevant states let
us first consider a {\it generic} pure state in  the Hilbert space
in $\mathbb{C}^{m}\ot \mathbb{C}^{n}$ ($m\leq n$). Such a generic
state (normalized, i.e. unit vector) has the form
\begin{equation}
|\Psi\rangle=\sum_{i=1}^m\sum_{j=1}^n
\alpha_{ij}|i\rangle|j\rangle,
\end{equation}
where the complex numbers $\alpha_{ij}$ may be regarded as random
variables distributed uniformly on a hypersphere, i.e. distributed
according to the probability density
\begin{equation}
P(\alpha)\propto \delta\left(\sum_{i=1}^m\sum_{j=1}^n
|\alpha_{ij}|^2-1\right), \label{distalpha}
\end{equation}
with the only constraint being the normalization. As we shall see,
such a generic state fulfills on average a ``volume" rather than
an area law. To this aim we introduce a somewhat more rigorous
description, and we prove that on average, the entropy of one of
subsystems of bipartite pure states in $\mathbb{C}^{m}\ot
\mathbb{C}^{n}$ ($m\leq n$) is almost maximal for sufficiently
large $n$. In other words, typical pure states in
$\mathbb{C}^{m}\ot \mathbb{C}^{n}$ are almost maximally entangled.
This ``typical behavior" of pure states happens to be completely
atypical for ground states of local Hamiltonians with an energy
gap between ground and first excited eigenstates.

Rigorously speaking, the average with respect to the distribution
(\ref{distalpha}) should be taken with respect to the unitarily
invariant measure on the projective space $\mathbb{C}P^{mn-1}$. It
is a unique measure generated by the Haar measure on the unitary
group by applying the unitary group on an arbitrarily chosen pure
state. One can show then that the eigenvalues of the first subsystem
of a randomly generated pure state $\ket{\psi_{AB}}$ are distributed
according to the following probability distribution
\cite{Lubkin78,LloydPagels88,Page93} (see also Ref.\ \cite{BengtssonZyczkowski}):

\begin{equation}\label{ProbDistr}
P_{m,n}(\lambda_{1},\ldots,\lambda_{m})=C_{m,n}
\delta\big(\sum_{i}\lambda_{i}-1\big)\prod_{i}\lambda_{i}^{n-m}\prod_{i<j}(\lambda_{i}-\lambda_{j})^{2},
\end{equation}
where the delta function is responsible for the normalization, and
the normalization constant reads (see e.g. Ref.\ \cite{BengtssonZyczkowski})
\begin{equation}
C_{m,n}=\frac{\Gamma(mn)}{\prod_{i=0}^{m-1}\Gamma(n-i)\Gamma(m-i+1)}
\end{equation}
with $\Gamma$ being the Euler gamma function\footnote{In general
the gamma function is defined through
\begin{equation}
\Gamma(z)=\int_{0}^{\infty}t^{z-1}e^{-t}\mathrm{d}t \qquad
(z\in\mathbb{C}).
\end{equation}
For $z$ being positive integers $z=n$ the gamma function is
related to the factorial function via $\Gamma(n)=(n-1)!$}.

\begin{theorem}\label{typical}
Let $\ket{\psi_{AB}}$ be a bipartite pure state from
$\mathbb{C}^{m}\ot\mathbb{C}^{n}$ $(m\leq n)$ drawn at random
according to the Haar measure on the unitary group and
$\varrho_{A}=\tr_{B}\proj{\psi_{AB}}$ be its subsystem acting on
$\mathbb{C}^{m}$. Then,
\begin{equation}\label{EntrAppr0}
\langle S(\varrho_{A})\rangle \thickapprox \log m-\frac{m}{2n}.
\end{equation}
\end{theorem}
\begin{proof}
Let us give here just an intuitive proof without detailed
mathematical discussion (which can be found e.g. in Refs.
\cite{Lubkin78,LloydPagels88,Page93,Foong94,Sen96,Sanchez--Ruiz95,BengtssonZyczkowski}).

Our aim is to estimate the following quantity
\begin{equation}\label{MeanEntropy}
\langle
S(\varrho_{A})\rangle=-\int\left(\sum_{i=1}^{m}\lambda_{i}\log
\lambda_{i}\right)P(\lambda_{1},\ldots,\lambda_{m})\,\mathrm{d}\lambda_{1}\ldots\mathrm{d}\lambda_{1},
\end{equation}
where the probability distribution
$P(\lambda_{1},\ldots,\lambda_{m})$ is given by Eq.
(\ref{ProbDistr}). We can always write the eigenvalues
$\lambda_{i}$ as $\lambda_{i}=\frac{1}{m}+\delta_{i}$, where
$\delta_{i}\in\mathbb{R}$ and $\sum_{i}\delta_{i}=0$. This allows
us to expand the logarithm into the Taylor series in the
neighborhood of $1/m$ as
\begin{eqnarray}\label{LogTaylor}
\log\left(\frac{1}{m}+\delta_{i}\right)=
-\log m+\sum_{k=1}^{\infty}\frac{(-1)^{k-1}}{k}(m\delta_{i})^{k},
\end{eqnarray}
which after application to Eq.\ (\ref{MeanEntropy}) gives the
following expression for the mean entropy
\begin{equation}
\langle S(\varrho_{A})\rangle=\log m-\frac{m}{1\cdot
2}\left\langle\sum_{i}\delta_{i}^{2}\right\rangle
+\frac{m^{2}}{2\cdot
3}\left\langle\sum_{i}\delta_{i}^{3}\right\rangle-
\frac{m^{3}}{3\cdot
4}\left\langle\sum_{i}\delta_{i}^{4}\right\rangle-\ldots.
\end{equation}
Let us now notice that
$\tr\varrho_{A}^{2}=\sum_{i}\lambda_{i}^{2}=\sum_{i}(\delta_{i}+1/m)^{2}=\sum_{i}\delta_{i}^{2}+1/m$,
and therefore $\sum_{i}\delta_{i}^{2}=\tr\varrho_{A}^{2}-1/m$.
This, after substitution in the above expression, together with the fact that
for sufficiently large $n$ we can omit terms with higher
powers of $\delta_{i}$ (cf. \cite{Lubkin78}), leads us to
\begin{equation}\label{EntropyAppr}
\langle S(\varrho_{A})\rangle\thickapprox \log
m-\frac{m}{2}\left\langle\tr\varrho_{A}^{2}-\frac{1}{m}\right\rangle.
\end{equation}
One knows that $\tr\varrho_{A}^{2}$ denotes the purity of
$\varrho_{A}$. Its average was calculated by Lubkin \cite{Lubkin78}
and reads
\begin{equation}\label{Trrho2}
\left\langle\tr\varrho_{A}^{2}\right\rangle=\frac{m+n}{mn+1}.
\end{equation}
Substitution in Eq.\ (\ref{EntropyAppr}) leads to the desired
results, completing the proof. $\blacksquare$

Two remarks should be made before discussing the area laws. First,
it should be pointed out that it is possible to get analytically
the exact value of $\langle S\rangle$. There is a series of papers
\cite{Foong94,Sen96,Sanchez--Ruiz95} presenting different
approaches leading to
\begin{equation}\label{EntrForm}
\langle S(\varrho)\rangle=\Psi(mn+1)-\Psi(n+1)-\frac{m-1}{2n}
\end{equation}
with $\Psi$ being the bigamma function\footnote{The bigamma function is
defined as $\Psi(z)=\Gamma'(z)/\Gamma(z)$ and for natural $z=n$ it
takes the form
\begin{equation}
\Psi(n)=-\gamma+\sum_{k=1}^{n-1}\frac{1}{n}
\end{equation}
with $\gamma$ being the Euler constant, of which exact value is
not necessary for our consideration as it vanishes in Eq.\
(\ref{EntrForm}).}. Using now the fact that
$\Psi(z+1)=\Psi(z)+1/z$, and the asymptotic properties of bigamma
function, $\Psi(z)\approx \log z$, we get (\ref{EntrAppr0}).

Second, notice that the exact result of Lubkin (\ref{Trrho2}) can
be estimated by relaxing the normalization constraint in the
distribution (\ref{distalpha}), and replacing it by a product of
independent Gaussian distributions,
$P(\alpha)=\prod_{i,j}(nm/\pi)\exp[-nm|\alpha_{ij}|^2]$, with
$\langle \alpha_{ij}\rangle=0$, and $\langle
|\alpha_{ij}|^2\rangle=1/nm$. The latter distribution, according
to the central limit theorem, tends for $nm\to \infty$ to a
Gaussian distribution for $\sum_{i=1}^m\sum_{j=1}^n
|\alpha_{ij}|^2$ centered at 1, with width $\simeq 1/\sqrt{nm}$.
One obtains then straightforwardly $\langle {\rm
tr}\varrho_A\rangle=1$, and after a little more tedious
calculation $\langle {\rm tr}\varrho^2_A\rangle=(n+m)/nm$, which
agrees asymptotically with the Lubkin result for $nm\gg 1$.

\end{proof}

\subsection{Area laws in a nutshell}

\begin{figure}[t!] 
   \centering
   \includegraphics[width=0.70\textwidth]{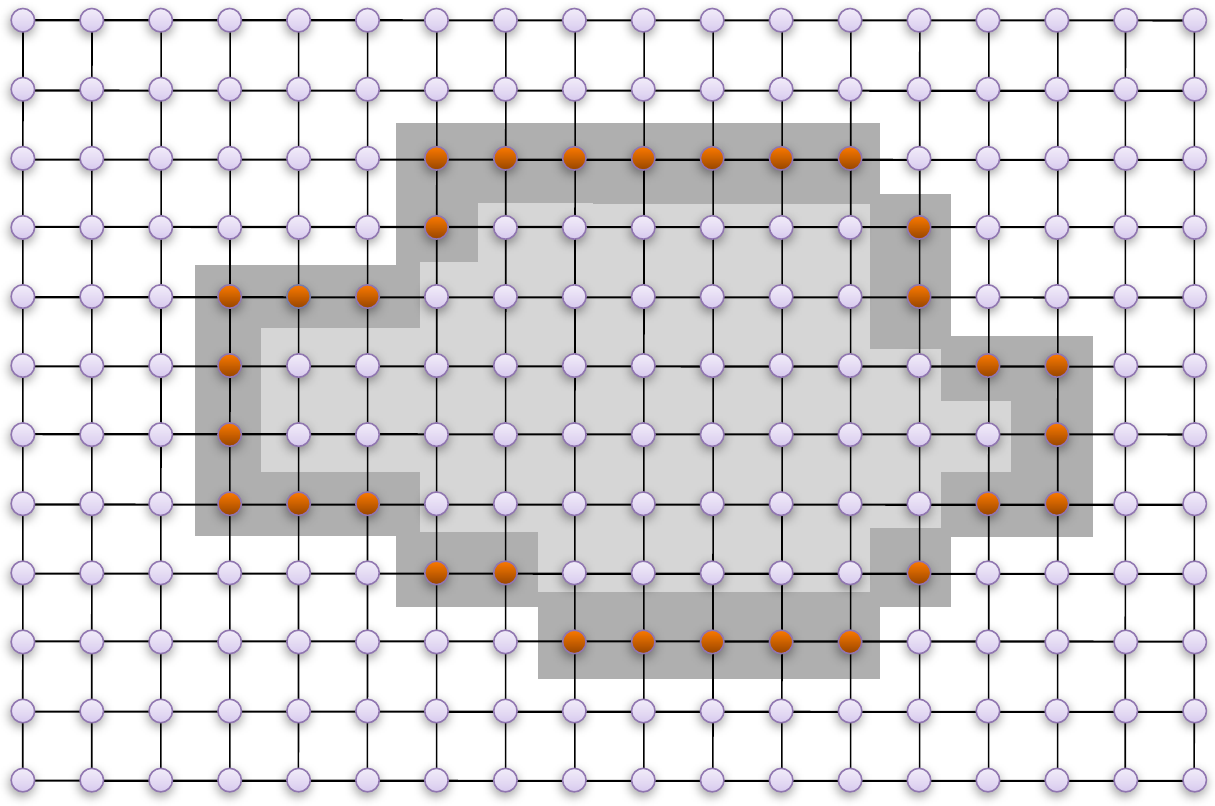}
   \caption{Schematic representation of a lattice system $L$, an arbitrary region
   $R$ (denoted in light grey background), and its boundary $\partial R$ (denoted
   in dark grey background).}
   \label{FigureAreaLaw}
\end{figure}

In what follows we shall be concerned with lattices $L$ in $D$
spatial dimensions, $L\subseteq \mathbb{Z}^{D}$. At each site we
have a $d$-dimensional physical quantum system (one can, however,
consider also classical lattices, with a $d$-dimensional classical
spin at each site with the configuration space
$\mathbb{Z}_{d}=\{0,\ldots,d-1\}$) at each site\footnote{For
results concerning other kind of systems one can consult Ref.\
\cite{EisertCramerPlenioReview}.}. The distance between two sites
$x$ and $y$ of the lattice is defined as
\begin{equation}
\mathcal{D}(x,y)=\max_{1\leq i\leq D}|x_{i}-y_{i}|.
\end{equation}
Accordingly, we define the distance between two disjoint
regions $X$ and $Y$ of $L$ as the minimal distance between all
pairs of sites $\{x,y\}$, where $x\in X$ and $y\in Y$; i.e.,
$\mathcal{D}(X,Y)=\min_{x\in X}\min_{y\in Y}\mathcal{D}(x,y)$.
If $R$ is some region of $L$, we define its boundary $\partial R$ as the set of sites
 belonging to $R$ whose distance to $L\setminus R$
(the complement of $R$) is one. Formally,
$\partial R=\{x\in R|\mathcal{D}(x,L\setminus R)=1\}$.
Finally, by $|R|$ we denote number of sites (or volume) in the region $R$
(see Figure \ref{FigureAreaLaw}).

Now, we can add some physics to our lattice by assuming that
interactions between the sites of $L$ are governed by some
hamiltonian $H$. We can divide the lattice $L$ into two parts, the
region $R$ and its complement $L\setminus R$. Roughly speaking, we
aim to understand how the entropy of the subsystem $R$ scales with
its size. In particular, we are interested in the entropy of the
state $\varrho_{R}$ reduced from a ground state or a thermal state
of the Hamiltonian $H$. We say that the entropy satisfies an area
law if it scales at most as the boundary area\footnote{Let us
shortly recall that the notation $f(x)=O(g(x))$ means that there
exist a positive constant $c$ and $x_{0}>0$ such that for any
$x\geq x_{0}$ it holds that $f(x)\leq cg(x)$.}, i.e.,
\begin{equation}\label{AreaLawGen}
S(\varrho_{R})=O(|\partial R|).
\end{equation}

\subsubsection{One-dimensional systems}

Let us start with the simplest case of one-dimensional
lattices,  $L=\{1,\ldots,N\}$. Let $R$ be a subset of $L$
consisting of $n$ contiguous spins starting from the first site,
i.e., $R=\{1,\ldots,n\}$ with $n<N$. In this case the boundary
$\partial R$ of the region $R$ contains one spin for open boundary conditions, and
two for periodic ones. Therefore, in this case the area law is extremely simple:
\begin{equation}\label{1DAreaLaw}
S(\varrho_{R})=O(1).
\end{equation}

The case of $D=1$ seems to be quite well understood. In general,
all local gapped systems (away from criticality) satisfy the above
law, and there might be a logarithmic divergence of entanglement
entropy when the system is critical. To be more precise, let us
recall the theorem of Hastings leading to the first of the above
statements, followed by examples of critical systems showing a
logarithmic divergence of the entropy with the size
of $R$.

Consider the nearest-neighbor interaction Hamiltonian
\begin{equation}\label{OneDHamilt}
H=\sum_{i\in L}H_{i,i+1},
\end{equation}
where each $H_{i,i+1}$ has a nontrivial support only on the sites
$i$ and $i+1$. We assume also that the operator norm of all the
terms in Eq. (\ref{OneDHamilt}) are upper bounded by some positive
constant $J$, i.e., $\|H_{i,i+1}\|\leq J$ for all $i$ (i.e., we
assume that the interaction strength between $i$th site and its
nearest-neighbor is not greater that some constant). Under these
assumptions, Hastings proved the following \cite{HastingsJSM}:
\begin{theorem}
Let $L$ be a one-dimensional lattice with $N$ $d$-dimensional sites,
and let $H$ be a local Hamiltonian as in Eq.\ (\ref{OneDHamilt}).
Assuming that $H$ has a unique ground state separated from the
first excited states by the energy gap $\Delta E>0$, the entropy
of any region $R$ satisfies
\begin{equation}
S(\varrho_{R})\leq 6c_{0}\xi2^{6\xi\log d}\log\xi\log d
\end{equation}
with $c_{0}$ denoting some constant of the order of unity and
$\xi=\min\{2v/\Delta E,\xi_{C}\}$. Here, $v$ denotes the sound
velocity and is of the order of $J$, while $\xi_{C}$ is a length scale of
order unity.
\end{theorem}

Let us remark that both constants appearing in the above theorem
come from the Lieb-Robinson bound \cite{LiebRobinson} (see also
Ref.\ \cite{Masanes} for a recent simple proof of this bound).

This theorem tells us that when the one-dimensional system with
the local interaction defined by Eq.\ (\ref{OneDHamilt}) is away
from the criticality $(\Delta E>0)$, the entropy of $R$ is bounded
by some constant independent of $|R|$ --- even if this bound does
not have to be tight. Of course, we can naturally ask if there
exist gapped systems with long-range interaction violating
(\ref{1DAreaLaw}). This was answered in the affirmative in Ref.
\cite{Dur05,EisertOsborne}, which gave examples of one-dimensional
models with long--range interactions, nonzero energy gap, and
scaling of entropy diverging logaritmically with $n$.

The second question one could pose is about the behavior of the
entropy when the gap $\Delta E$ goes to zero and the system
becomes critical. Numerous analytical and numerical results show
that usually one observes a logarithmic divergence of
$S(\varrho_{R})$ with the size of the region $R$. Here we recall
only the results obtained for the so--called XY model in a
transverse magnetic field (for the remaining ones we refer the reader to recent reviews
\cite{EisertCramerPlenioReview,Latorre09}, and to the special issue of
J. Phys. A devoted to this subject \cite{Calabrese10}).

The Hamiltonian for the XY model reads
\begin{equation}\label{XYHamil}
H_{XY}=-\frac{1}{2}\sum_{ i\in
L}\left(\frac{1+\gamma}{2}\sigma_{i}^{x}
\sigma_{i+1}^{x}+\frac{1-\gamma}{2}\sigma_{i}^{y}\sigma_{i+1}^{y}\right)-\frac{h}{2}\sum_{i\in
L}\sigma_{i}^{z},
\end{equation}
with $0\leq \gamma\leq 1$ the anisotropy parameter, and $h$ the
magnetic field. In the case of vanishing anisotropy
($\gamma=0$), we have the isotropic XY model called shortly XX
model, while for $\gamma=1$ one recovers the well--known Ising
Hamiltonian in a transverse field. The Hamiltonian $H_{XY}$ is
critical when either $\gamma=0$ and $|h|\leq 1$ (the critical XX
model) or for $|h|=1$.

It was shown in a series of papers
\cite{Vidal03b,JinKorepin04,Its05,KeatingMezzadri05} that for the
critical XY model (that is when $\gamma\neq 0$ and $|h|= 1$) the
entropy of the region $R=\{1,\ldots,n\}$ scales as
\begin{equation}
S(\varrho_{R})=\frac{1}{6}\log_{2}n+O(1),
\end{equation}
while for the critical XX model, the constant multiplying the
logarithms becomes one--third. Then, in the case of the critical
Ising model ($\gamma=1$), it can be shown that the entropy scales
at least logaritmically\footnote{The notation $f(x)=\Omega(g(x))$
means that there exist $c>0$ and $x_{0}>0$ such that $f(x)\geq
cg(x)$ for all $x\geq x_{0}$.}, i.e.,
$S(\varrho_{R})=\Omega(\log_{2}n)$
\cite{EisertCramerPlenioReview,Eisert05}.

Concluding, let us mention that there is an extensive literature
on the logarithmic scaling of the block entropy using conformal
field theory methods (see Ref.\ \cite{Calabrese09} for a very good
overview of these results). Quite
generally, the block entropy at criticality scales as
\begin{equation}
S(\varrho_{R})=\frac{c}{3}\log_{2}\left(\frac{|R|}{a}\right)+O(1),
\end{equation}
or, more in general for the R\'enyi entropy\footnote{Recall that
the quantum R\'enyi entropy is defined as
\begin{equation}
S_{\alpha}=\frac{1}{1-\alpha}\log_{2}\left[\Tr\left(\varrho^{\alpha}\right)\right]
\end{equation}
where $\alpha\in[0,\infty]$. For $\alpha=0$ one has
$S_{0}(\varrho)=\log_{2}\mathrm{rank}(\varrho)$ and
$S_{\infty}=-\log_{2}\lambda_{\mathrm{max}}$ with
$\lambda_{\mathrm{max}}$ being the maximal eigenvalue of
$\varrho$. }
\be
S_{\alpha}(\varrho_{R})=(c/6)\left(1+1/\alpha\right)\log_{2}(|R|/a)+O(1),
\ee
where $c$ is called the {\it central charge} of the
underlying conformal field theory, and $a$ is the cutoff parameter
(the lattice constant for lattice systems).

\subsubsection{Higher--dimensional systems}

The situation is much more complex in higher spatial dimensions
$(D>1)$. The boundary $\partial R$ of the general area law, Eq.
(\ref{AreaLawGen}), is no longer a simple one or two--element set
and can have a rather complicated structure. Even if there are no
general rules discovered so far, it is rather believed that
(\ref{AreaLawGen}) holds for ground states of local gapped
Hamiltonians. This intuition is supported by results showing that
for quadratic quasifree fermionic and bosonic lattices the area
law (\ref{AreaLawGen}) holds \cite{EisertCramerPlenioReview}.
Furthermore, for critical fermions the entropy of a cubic region
$R=\{1,\ldots,n\}^{D}$ is bounded  as $\gamma_1
n^{D-1}\log_{2}n\leq S(\varrho_{R})\leq \gamma_2
n^{D-1}(\log_{2}n)^{2}$ with $\gamma_{i}$ $(i=1,2)$ denoting some
constants  \cite{WolfPRL06,GioevKlich06,FarkasZimboras07}. Let us
notice that the proof of this relies on the fact that logarithmic
negativity (see Eq.\ (\ref{LogNeg})) upper bounds the von Neumann
entropy, i.e., for any pure bipartite state $\ket{\psi_{AB}}$, the
inequality $S(\varrho_{A(B)})\leq E_{N}(\ket{\psi_{AB}})$ holds.
This in turn is a consequence of monotonicity of the R\'enyi
entropy $S_{\alpha}$ with respect to the order $\alpha$, i.e.,
$S_{\alpha}\leq S_{\alpha'}$ for $\alpha\geq \alpha'$. This is one
of the instances where insights from quantum information help to
deal with problems in many--body physics.

Interestingly, very recently Masanes \cite{Masanes} showed that
the ground state (and also low--energy eigenstates) entropy of a
region $R$ (even a disjoint one) always scales at most as the size
of the boundary of $R$ with some correction proportional to
$(\log|R|)^{D}$ --- as long as the Hamiltonian $H$ is of the local
form
\begin{equation}
H=\sum_{i\in L}H_{i},
\end{equation}
where each $H_{i}$ has nontrivial support only on the
nearest-neighbors of the $i$th site, and satisfies as previously
$\|H_{i}\|\leq J$ for some $J>0$. Thus, the behavior of entropy
which is considered to be a violation of the area law
(\ref{AreaLawGen}) can in fact be treated as an area law itself.
This is because in this case\footnote{It should be noticed that
one can have much stronger condition for such scaling of entropy.
To see this explicitly, say that $R$ is a cubic region
$R=\{1,\ldots,n\}^{D}$ meaning that $|\partial R|=n^{D-1}$ and
$|R|=n^{D}$. Then since $\lim_{n\to\infty}[(\log
n)/n^{\epsilon}]=0$ for any (even arbitrarily small) $\epsilon>0$,
one easily checks that $S(\varrho_{R})/|\partial
R|^{1+\epsilon}\to 0$ for $|\partial R|\to \infty$.}
$[|\partial R|(\log|R|)^{k}]/|R|\to 0 $ for $|R|\to \infty$ with
some $k>0$, meaning that still this behavior of entropy is very
different from the typical behavior following from theorem
\ref{typical}. That is, putting $m=d^{|R|}$ and $n=d^{|L\setminus
R|}$ with $|L|\gg|R|$ one has that $S(\varrho_{R})/|R|$ is
arbitrarily close to $\log d$ for large $|R|$.

Let $R_{1}$ and $R_{2}$ be two disjoint regions of the lattice
such that $|R_{1}|\leq |R_{2}|$, and let $l$ denote the distance
between these regions. Let us call $\Gamma$ a function that bounds
from above the correlations between two operators $X$ and $Y$
($\|X\|,\|Y\|\leq 1$) acting respectively on $R_{1}$ and $R_{2}$,
i.e., $C(X,Y)=|\langle XY\rangle-\langle X\rangle\langle
Y\rangle|\leq \Gamma(l,|R_{1}|)$. The first assumption leading to
the results of Ref.\ \cite{Masanes} is that if the mean values in
$C$ are taken in the ground state of $H$, $\Gamma$ is given by
\begin{equation}\label{as1}
\Gamma(l,|R_{1}|)=c_{1}(l-\xi\log|R_{1}|)^{-\mu}
\end{equation}
with some constants $c_{1}$, $\xi$, and $\mu>D$. Notice that this
function decays polynomially in $l$, meaning that this first
assumption is weaker than the property of exponential decay
observed in Ref.\ \cite{Hastings04} for gapped Hamiltonians.

Let now $H_{R}$ denote a part of the global Hamiltonian $H$ which
acts only on sites in some region $R$. It has its own eigenvalues
and eigenstates, denoted by $e_{n}$ and $\ket{\psi_{n}}$
respectively, with $e_{0}$ denoting the lowest eigenvalue. The
second assumption made in Ref.\ \cite{Masanes} is that there exist
constants $c_{2}$, $\tau$, $\gamma$, and $\eta$ such that for any
region $R$ and energy $e=2J3^{D}|\partial R|+e_{0}+40v$ (notice that $3^D$ is
the number of the first neighbours in a cubic lattice), the
number of eigenenergies of $H_{R}$ lower than $e$ is upper bounded
as
\begin{equation}\label{BoundEnergies}
\Omega_{R}(e)\leq c_{2}(\tau|R|)^{\gamma(e-e_{0})+\eta|\partial
R|}.
\end{equation}
Now, we are in position to formulate the main result of Ref.
\cite{Masanes}.

\begin{theorem}
Let $R$ be some arbitrary (even disjoint) region of $L$. Then,
provided the assumptions (\ref{as1}) and (\ref{BoundEnergies})
hold, the entropy of the reduced density matrix $\varrho_{R}$ of
the ground state of $H$ satisfies
\begin{equation}\label{formula}
S(\varrho_{R})\leq C|\partial R|(10\xi\log|R|)^{D}
+O(|\partial R|(\log|R|)^{D-1}),
\end{equation}
\end{theorem}
where $C$ collects the constants $D,\xi,\gamma,J,\eta$, and $d$.
If $R$ is a cubic region, the above statement simplifies, giving
$S(\varrho_{R})\leq \widetilde{C}|\partial R|\log|R|+O(|\partial
R|)$ with $\widetilde{C}$ being some constant.

Leaving out the first assumption, however, at the cost of extending the
second assumption to all energies $e$ (not only the ones bounded
by $2J3^{D}|\partial R|+e_{0}+40v$), leads to the following simple
area law:
\begin{theorem}
Let $R$ be an arbitrary region of the lattice $L$. Assuming that
the above number of eigenvalues $\Omega_{R}(e)$ satisfies
condition (\ref{BoundEnergies}) for all $e$, then
\begin{equation}\label{SimpleAreaLaw}
S(\varrho_{R})\leq C|\partial R|\log|R|+O(|\partial R|).
\end{equation}
\end{theorem}
\begin{proof}
Let $\ket{\psi_{i}}$ and $e_{i}$ denote the eigenvectors and
ordered eigenvalues $(e_{0}\leq e_{1}\leq \ldots \leq e_{n}\leq
\ldots)$ of $H_{R}$. Then, it is clear that the ground state
$\ket{\Psi_{0}}$ of $H$ can be written as
$\ket{\Psi_{0}}=\sum_{i,j}\alpha_{ij}\ket{\psi_{i}}\ket{\varphi_{j}}$,
where the vectors $\ket{\varphi_{j}}$ constitute some basis in the
Hilbert space corresponding to the region $L\setminus R$. One may
always denote
$\sqrt{\mu_{i}}\ket{\widetilde{\varphi}_{i}}=\sum_{j}\alpha_{ij}\ket{\varphi_{j}}$,
and then
\begin{equation}\label{FormPhiZero}
\ket{\Psi_{0}}=\sum_{i}\sqrt{\mu_{i}}\,\ket{\psi_{i}}\ket{\widetilde{\varphi}_{i}},
\end{equation}
where
$\mu_{i}=1/\langle\widetilde{\varphi}_{i}|\widetilde{\varphi}_{i}\rangle
=1/\sum_{j}|\alpha_{ij}|^{2}\geq 0$ and they add up to unity. The
vectors $\ket{\widetilde{\varphi}_{i}}$ in general do not have to
be orthogonal, therefore Eq.\ (\ref{FormPhiZero}) should not the
confused with the Schmidt decomposition of $\ket{\Psi_{0}}$.
Nevertheless, one may show that tracing out the
$L\setminus R$ subsystem the entropy of the density
matrix acting on $R$ is upper bounded as (see Ref.\ \cite{Bennett96a})
\begin{equation}\label{EntrIneq}
S(\varrho_{R})\leq -\sum_{i}\mu_{i}\log\mu_{i}.
\end{equation}
We now aim to maximize the right-hand side of the above equation
under the following conditions imposed on
$\mu_{i}$: First, the locality of our
Hamiltonian means that $\langle H_{R}\rangle\leq
e_{0}+J3^{D}|\partial R|$, implying that the probabilities $\mu_{i}$
obey
\begin{equation}\label{MeanEnergy}
\sum_{i}\mu_{i}\widetilde{e}_{i}\leq J3^{D}|\partial R|,
\end{equation}
with $\widetilde{e}_{i}=e_{i}-e_{0}$. Second, the modified version
of the second assumption allows to infer that for any eigenvalues
$e_{i}$ the inequality $i\leq
c_{2}(\tau|R|)^{\gamma\widetilde{e}_{i}+\eta|\partial R|}$ holds.
Substitution of the above in Eq.\ (\ref{MeanEnergy}) gives
\begin{equation}\label{Condition}
\sum_{i}\mu_{i}\log i\leq C|\partial R|\log|R|+O(|\partial R|)
\end{equation}
where $C$ contains the constants $\eta,\gamma,J$, and $D$.
Eventually, following the standard convex optimization method (see
e.g. Ref.\ \cite{BoydVanderberghe}) with two constraints
(normalization and the inequality (\ref{Condition})) one gets
(\ref{SimpleAreaLaw}). $\blacksquare$

\end{proof}

\subsubsection{Are laws for mutual information - classical and quantum Gibbs states}

So far, we considered area laws only for ground states of local
Hamiltonians. In addition, it would be very interesting to ask
similar questions for nonzero temperatures. Here, however, one
cannot rely on the entropy of a subsystem, as in the case of mixed
states it looses its meaning. A very good quantity measuring the
total amount of correlation in bipartite quantum systems is the
{\it quantum mutual information} \cite{GroismanPopescuWinterPRA05}
defined as
\begin{equation}\label{Mutual}
I(A:B)=S(\varrho_{A})+S(\varrho_{B})-S(\varrho_{AB}),
\end{equation}
where $\varrho_{AB}$ is some bipartite state with its subsystems
$\varrho_{A(B)}$. It should be noticed that for pure states the
mutual information reduces to twice the amount of entanglement of
the state.

Recently, it was proven that thermal states
$\varrho_{\beta}=e^{-\beta H}/\tr[e^{-\beta H}]$ with local
Hamiltonians $H$ obey an area law for mutual information.
Interestingly, a similar conclusion was drawn for classical
lattices, in which at each site we have a classical spin with the
configuration space $\mathbb{Z}_{d}$, and instead of density
matrices one deals with probability distributions. In the following
we review these two results, starting from the classical
case.

To quantify correlations in classical systems, we use the
classical mutual information,  defined as in Eq.\ (\ref{Mutual})
with the von Neumann entropy substituted by the Shannon entropy
$H(X)=-\sum_{x}p(x)\log_2 p(x)$, where $p$ stands for a
probability distribution characterizing random variable $X$. More
precisely, let $A$ and $B=S\setminus A$ denote two subsystems of
some classical physical system $S$. Then, let $p(x_A)$ and
$p(x_B)$ be the marginals of the joint probability distribution
$p(x_{AB})$ describing $S$ ($x_{a}$ denotes the possible
configurations of subsystems $a=A,B,AB$). The correlations between
$A$ and $B$ are given by
\begin{equation}\label{MutualInformation}
I(A:B)=H(A)+H(B)-H(AB).
\end{equation}
We are now ready to formulate and prove the following theorem
\cite{Wolf08}.
\begin{theorem}
Let $L$ be a lattice with $d$--dimensional classical spins at each
site. Let $p$ be a Gibbs probability distribution coming from
finite--range interactions on $L$. Then, dividing $L$ into regions
$A$ and $B$, one has
\begin{equation}\label{AreaLawClassicalMutInf}
I(A:B)\leq |\partial A|\log d.
\end{equation}
\end{theorem}
\begin{proof}
First, notice that the Gibbs distributions coming from
finite--range interactions have the property that if a region $C$
separates $A$ from $B$ in the sense that no interaction is between
$A$ and $B$ then $p(x_{A}|x_{C},x_{B})=p(x_{A}|x_{C})$, which we
rewrite as
\begin{equation}\label{MarkovProb}
p(x_{A},x_{B},x_{C})=\frac{p(x_{A},x_{C})p(x_{B},x_{C})}{p(x_{C})}.
\end{equation}
Now, let $A$ and $B$ be two regions of $L$, and let $\partial
A\subset A$ and $\partial B\subset B$ be boundaries of $A$ and
$B$, respectively, collecting all sites interacting with their
exteriors. Finally, let $\overline{A}=A\setminus\partial A$ and
$\overline{B}=B\setminus\partial B$. Since $\partial A$ separates
$A$ from $\partial B$ (there is no interaction between $A$ and
$\partial B$), we can use Eq.\ (\ref{MarkovProb}) to obtain
\begin{eqnarray}\label{EntrDerivation}
H(AB)&=&H(\bar{A}\partial AB)=-\sum_{x_{\bar{A}},x_{\partial
A},x_{B}}p(x_{\bar{A}},x_{\partial A},x_{B})\log_2
p(x_{\bar{A}},x_{\partial
A},x_{B})\nonumber\\
&=&-\sum_{x_{\bar{A}},x_{\partial A}}p(x_{\bar{A}},x_{\partial
A})\log_2 p(x_{\bar{A}},x_{\partial
A})\nonumber\\
&&-\sum_{x_{\partial A},x_{B}}p(x_{\partial A},x_{B})\log_2
p(x_{\partial
A},x_{B})\nonumber\\
&&+\sum_{x_{\partial A}}p(x_{\partial A})\log_2 p(x_{\partial
A})\nonumber\\
&=&H(A)+H(\partial AB)-H(\partial A).
\end{eqnarray}
Since $\partial B$ separates $\partial A$ from $B$, the same
reasoning may be applied to the second term of the right-hand side
of the above, obtaining $H(\partial AB)=H(\partial A\partial
B)+H(B)-H(\partial B)$. This, together with Eq.
(\ref{EntrDerivation}), gives
\begin{equation}
H(AB)=H(A)+H(B)+H(\partial A\partial B)-H(\partial A)-H(\partial
B),
\end{equation}
which in turn after application to Eq.\ (\ref{MutualInformation})
allows us to write
\begin{equation}
I(A:B)=I(\partial A:\partial B).
\end{equation}
It means that whenever the probability distribution $p$ has the
above Markov property, correlations between $A$ and $B$ are the
same as between their boundaries.

Now, we know that the mutual information can be expressed through
the conditional Shannon entropy\footnote{The conditional Shannon
entropy is defined as $H(A|B)=H(A,B)-H(B)$.} as
$I(X:Y)=H(X)-H(X|Y)$. Since $H(X|Y)$ is always nonnegative, we
have the following inequality
\begin{equation}\label{ClassicalArea}
I(\partial A:\partial B)\leq H(\partial A)\log d.
\end{equation}
To get Eq.\ (\ref{AreaLawClassicalMutInf}) it suffices to notice
that $H(A)$ is upper bounded by the Shannon entropy of
independently and identically distributed probability
$p(x_{A})=1/d^{|A|}$, which means that $H(A)\leq |A|\log d$.
$\blacksquare$
\end{proof}
Let us now show that a similar conclusion can be drawn in the case
of quantum thermal states \cite{Wolf08}, where the Markov property
does not hold in general.
\begin{theorem}
Let $L$ be a lattice consisting of $d$-dimensional quantum systems
divided into parts $A$ and $B$ ($L=A\cup B$). Thermal states
$(T>0)$ of local Hamiltonians $H$ obey the following area law
\begin{equation}\label{MutualAreaLaw}
I(A:B)\leq \beta
\tr[H_{\partial}(\varrho_{A}\ot\varrho_{B}-\varrho_{AB})].
\end{equation}
\end{theorem}
\begin{proof}
The thermal state $\varrho_{\beta}=e^{-\beta H}/\tr(e^{-\beta H})$
 minimizes the free energy
$F(\varrho)=\tr(H\varrho)-(1/\beta)S(\varrho)$, and therefore
$F(\varrho_{\beta})\leq
F(\varrho^{A}_{\beta}\ot\varrho^{B}_{\beta})$ with
$\varrho^{A}_{\beta}$ and $\varrho^{B}_{\beta}$
subsystems of $\varrho_{\beta}$. This allows us to estimate the
entropy of the thermal state as
\begin{eqnarray}\label{QEntropyEst}
S(\varrho_{\beta})&=&\beta\left[\tr(H\varrho_{\beta})-
F(\varrho_{\beta})\right]\nonumber\\
&\geq&\beta\left[\tr(H\varrho_{\beta})-F(\varrho^{A}_{\beta}\ot\varrho^{B}_{\beta})\right]\nonumber\\
&=&\beta\left[\tr(H\varrho_{\beta})-\tr(H\varrho_{\beta}^{A}\ot\varrho_{\beta}^{B})\right]+S(\varrho_{\beta}^{A}\ot\varrho_{\beta}^{B})\nonumber\\
&=&\beta\left[\tr(H\varrho_{\beta})-\tr(H\varrho_{\beta}^{A}\ot\varrho_{\beta}^{B})\right]+S(\varrho_{\beta}^{A})+S(\varrho_{\beta}^{B}),
\end{eqnarray}
where the last equality follows from additivity of the von Neumann
entropy $S(\rho\ot\sigma)=S(\rho)+S(\sigma)$. Putting Eq.
(\ref{QEntropyEst}) into the formula for mutual information we get
\begin{equation}\label{QMutInfEst}
I(A:B)\leq
\beta\left[\tr(H\varrho_{\beta}^{A}\ot\varrho_{\beta}^{B})-\tr(H\varrho_{\beta})\right].
\end{equation}
Let us now write the Hamiltonian as $H=H_{A}+H_{\partial}+H_{B}$,
where $H_{A}$ and $H_{B}$ denote all the interaction terms within
the regions $A$ and $B$, respectively, while $H_{\partial}$ stands
for interaction terms connecting these two regions. Then one
immediately notices that
$\tr[H_{A(B)}(\varrho_{\beta}^{A}\ot\varrho_{\beta}^{B}-\varrho_{\beta})]=0$
and only the $H_{\partial}$ part of the Hamiltonian $H$
contributes to the right-hand side of Eq.\ (\ref{QMutInfEst}). This
finishes the proof. $\blacksquare$
\end{proof}
Let us notice that the right--hand side of Eq.
(\ref{MutualAreaLaw}) depends only on the boundary, and therefore
it gives a scaling of mutual information similar to the classical
case (\ref{ClassicalArea}). Moreover, for the nearest-neighbor
interaction, Eq.\ (\ref{MutualAreaLaw}) simplifies to $I(A:B)\leq
2\beta\|h\|\,|\partial A|$ with $\|h\|$ denoting the largest
eigenvalue of all terms of $H$ crossing the boundary.

\section{The tensor network product world}
\label{sec_10_4}

Quantum many-body systems are, in general, difficult to describe:
specifying an arbitrary state of a system with $N$-two level
subsystems requires $2^N$ complex numbers. For a classical
computer, this presents not only storage problems, but also
computational ones, since simple operations like calculating the
expectation value of an observable would require an exponential
number of operations. However, we know that completely separable
states can be described with about $N$ parameters --- indeed, they
correspond to classical states. Therefore, what makes a quantum
state difficult to describe are quantum correlations, or
entanglement. We saw already that even if in general the entropy
of a subsystem of an arbitrary state is proportional to the
volume, there are some special states which obey an entropic area
law. Intuitively, and given the close relation between entropy and
information, we could expect that states that follow an area law
can be described (at least approximately) with much less
information than a general state. We also know that such low
entanglement states are few, albeit interesting --- we only need
an efficient and practical way to describe and parameterize them.

\subsection{The tensor network representation of quantum states}
\begin{figure}[t!] 
   \centering
   \includegraphics[width=0.70\textwidth]{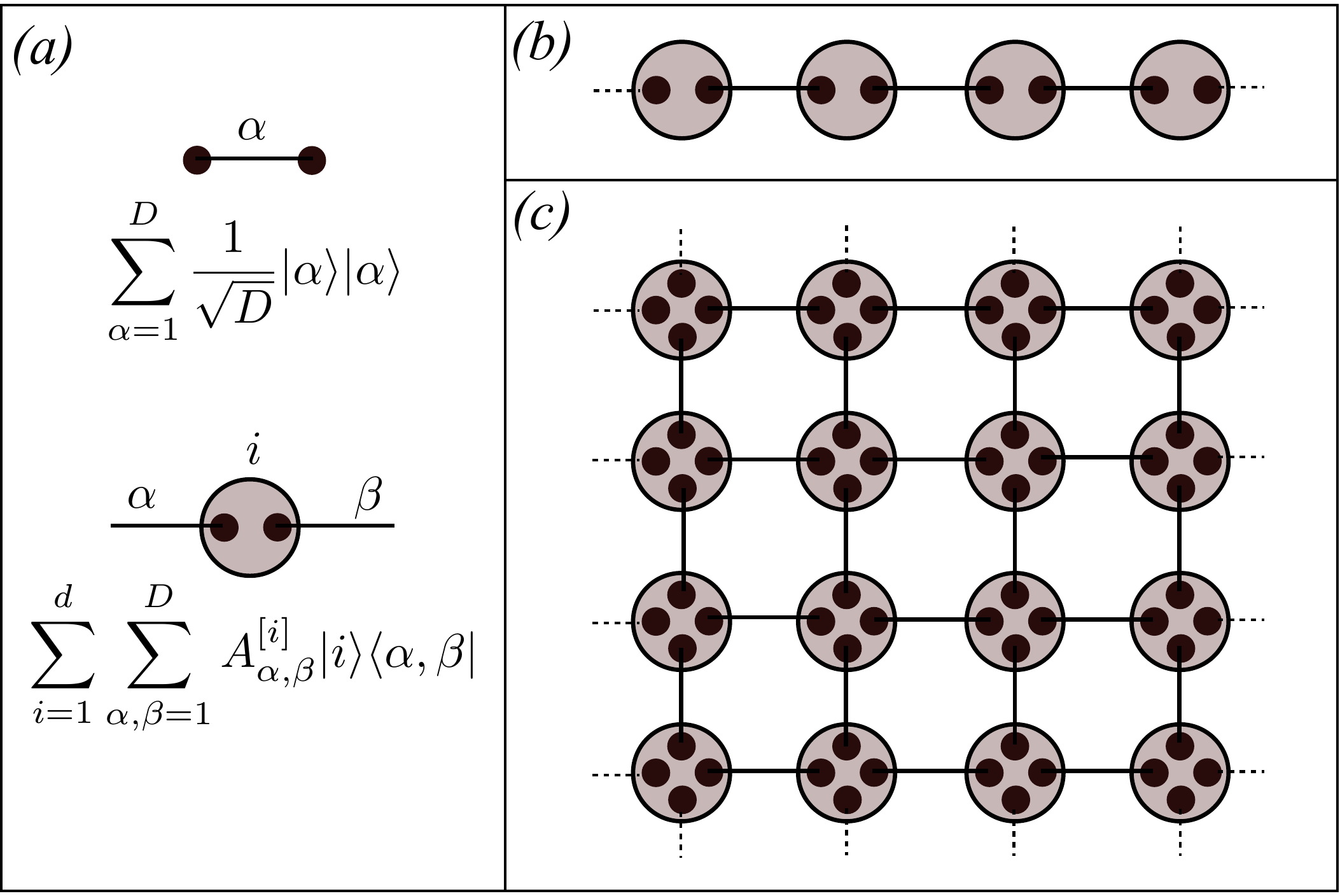}
   \caption{Schematic representation of tensor networks. In panel $(a)$ we show the meaning of the
   elements in the representation, namely the solid line joining two virtual particles in different sites
   means the maximally entangled state between them, and the grey circle represents the map
   from virtual particles in the same site to the physical index.
   In panel $(b)$  we see a one-dimensional tensor network or MPS, while in $(c)$ we show how
   the scheme can be extended intuitively to higher dimensions --- in the two-dimensional example
   shown here, a PEPS that contains four virtual particles per physical site.}
   \label{FigureMPS}
\end{figure}

Consider a general state of a system with $N$ $d$-level particles,
\begin{equation}
| \psi \rangle = \sum_{i_1,i_2,\ldots,i_N=1}^{d} c_{i_1i_2\ldots
i_N} | i_1,i_2,\ldots,i_N \rangle.
\end{equation}
When the state has no entanglement, then $c_{i_1i_2\ldots
i_N}=c^{(1)}_{i_1}c^{(2)}_{i_2}\ldots c^{(N)}_{i_N}$ where all
$c$'s are scalars. The locality of the information (the set of
coefficients $c$ for each site is independent of the others) is
key to the efficiency with which separable states can be
represented. How can we keep this locality while adding complexity
to the state, possibly in the form of correlations but only to
nearest-neighbors? As we shall see, we can do this by using a
tensor at each site of our lattice, with one index of the tensor
for every physical neighbor of the site, and another index for the
physical states of the particle. For example, in a one-dimensional
chain we would assign a matrix for each state of each particle,
and the full quantum state would write as
\begin{equation}
| \psi \rangle = \sum_{i_1,i_2,\ldots,i_N=1}^{d} {\rm tr} \left[
A^{[1]}_{i_1}A^{[2]}_{i_2}\ldots A^{[N]}_{i_N} \right] |
i_1,i_2,\ldots i_N \rangle, \label{mps}
\end{equation}
where $A^{[k]}_{i_k}$ stands for a matrix with dimensions $D_k
\times D_{k+1}$. A useful way of understanding the motivations for
this representation is to think of a valence bond
picture \cite{Verstraete04a}. Imagine that we replace every
particle at the lattice by a pair (or more in higher dimensions)
of particles of dimensions $D$ that are in a maximally entangled
state with their corresponding partners in a neighboring site (see
Figure \ref{FigureMPS}). Then, by applying a map from this virtual
particles into the real ones,
\begin{equation}
{\cal A}=\sum_{i=1}^d \sum_{\alpha,\beta=1}^D
A_{\alpha,\beta}^{[i]} |i\rangle\!\langle\alpha ,\beta |,
\end{equation}
we obtain a state that is expressed as Eq.\ (\ref{mps}). One can
show that any state $| \psi \rangle \in \mathbb{C}^{d N}$ can be
written in this way with $D=\max_m D_m \le d^{N/2}$. Furthermore,
a matrix product state can always be found such that
\cite{Vidal03a}
\begin{itemize}
\item $\sum_i A^{\dagger [k]}_{i} A^{[k]}_{i} = 1_{D_k}$, for $1 \le k \le N$,
\item $\sum_i A^{\dagger[k]}_{i} \Lambda^{[k-1]} A^{[k]}_{i}  = \Lambda^{[k]}$, for $1 \le k \le N$, and
\item For open boundary conditions $\Lambda^{[0]}=\Lambda^{[N]}=1$, and $\Lambda^{[k]}$ is a $D_{k+1} \times
D_{k+1}$ positive diagonal matrix, full rank, with ${\rm tr}
\Lambda^{[k]}=1$.
\end{itemize}
In fact, $\Lambda^{[k]}$ is a matrix whose diagonal components
$\lambda^{k}_n$, $n=1,\ldots,D_k$, are the non-zero eigenvalues of
the reduced density matrix obtained by tracing out the particles
from $k+1$ to $N$, i.e., the Schmidt coefficients of a bipartition
of the system at site $k$. A MPS with these properties is said to
be in its canonical form \cite{Perez-Garcia07}.

Therefore, Eq.\ (\ref{mps}) is a representation of all possible
states --- still cumbersome. It becomes an efficient
representation when the virtual bond dimension $D$ is small, in
which case it is typically said that the state has a matrix
product state (MPS) representation. In higher dimensions we talk
about projected entangled pair states (PEPS) \cite{Verstraete04d}.
When entanglement is small (but finite), most of the Schmidt
coefficients are either zero or decay rapidly to zero
\cite{Vidal03b}. Then, if $| \psi \rangle $ contains little
entanglement, we can obtain a very good approximation to it by
truncating the matrices $A$ to a rank $D$ much smaller than the
maximum allowed by the above theorem, $d^{N/2}$. In fact, we can
demonstrate the following
\begin{lemma} \cite{Perez-Garcia07}
 There exists a MPS $|\psi_D \rangle$ with bond dimension $D$ such that
$\| | \psi \rangle - | \psi_D \rangle \|^{2} < 2
\sum_{\alpha=1}^{N-1} \epsilon_\alpha(D)$, where
$\epsilon_\alpha(D) = \sum_{i=D+1}^{d^{\min(\alpha, N-\alpha)}}
\lambda^{[k]}_i$. \label{lemmaBound}
\end{lemma}
\begin{proof}
\smartqed Let us assume that the MPS is in its canonical form with
$D=2^{N/2}$. Defining a projector into the virtual bond dimension
$P=\sum_{k=1}^D |k\rangle\!\langle k|$, and a TPCM map $
\$_m(X)=\sum_i A^{[m]\dagger}_i X A^{[m]}_i$, we can write the
overlap
\begin{equation}
\langle \psi | \psi_D \rangle = {\rm Tr} \left[ \$_2(\ldots
\$_{N-2}(\$_{N-1}(\Lambda^{[N-1]} P) P) P) \ldots) P \right].
\end{equation}
By defining $Y^{[k]}=\$_k(Y^{[k+1]}P)$, with
$Y^{[N-1]}=\Lambda^{[N-1]} P$, and using that ${\rm Tr} | \$(X) |
\le {\rm Tr} |X|$, we can see that
\begin{eqnarray}
\tr \big| \Lambda^{[k]} - Y^{[k]} \big| &=& \tr \big|\$_k(
\Lambda^{[k+1]} - Y^{[k+1]}P) \big|\nonumber\\
&\le& \tr \big| \Lambda^{[k+1]} - Y^{[k+1]} \big| + \tr \big|
\Lambda^{[k+1]}(1-P) \big|,
\end{eqnarray}
where the last term  is equal to $\sum_{\alpha=D+1}^{2^{N/2}}
\lambda^{[k]}_\alpha$. Finally, applying this last inequality
recursively from $N-1$ to $2$, and using that $\langle \psi_D |
\psi_D \rangle \le 1$, we can obtain the desired bound on $\langle
\psi | \psi_D \rangle $. \qed
\end{proof}

Lemma (\ref{lemmaBound}) is most powerful in the context of
numerical simulations of quantum states: it gives a controllable
handle on the precision of the approximation by MPS. In practical
terms, for the representation to be efficient the Schmidt
coefficients $\lambda$ need to decay faster than polynomially.
However, we can be more precise and give bounds on the error of
the approximation in terms of entropies \cite{Schuch07}:
\begin{lemma}
Let $S_\alpha(\rho)=\log (\tr \rho^\alpha)/(1-\alpha)$ be the
R\'enyi entropy of a reduced density matrix $\rho$, with $0<
\alpha < 1$. Denote $\epsilon(D) = \sum_{i=D+1}^{\infty}
\lambda_i$, with $\lambda_i$ being the eigenvalues of $\rho$ in
nonincreasing order. Then,
\begin{equation}
\log (\epsilon(D)) \le \frac{1-\alpha}{\alpha} \left(
S_\alpha(\rho) - \log \frac{D}{1-\alpha} \right).
\end{equation}
\end{lemma}

The question now is when can we find systems with relevant states
that can be written efficiently as a MPS; i.e. how broad is the
simulability of quantum states by MPS. For example, one case of
interest where we could expect the method to fail is near quantum
critical points where correlations (and entanglement) are singular
and might diverge. However, at least in 1D systems, we can state
the following:
\begin{lemma} \cite{Perez-Garcia07}
In one dimension there exists a scalable, efficient MPS
representation of ground states even at criticality.
\end{lemma}
\begin{proof}
In one dimension, the worst case growth of entropy of a subsystem
of size $L$, exactly at criticality, is given by
\begin{equation}
S_\alpha(\rho_L)\simeq \frac{c+\tilde c}{12} \left(
1+\frac{1}{\alpha} \right) \log L.
\end{equation}
Let us take the length $L$ to be half the chain, $N=2L$. By means
of the previous discussion, we can find a MPS $|\psi_D\rangle$
such that its distance  with the ground state $| \psi_{GS}
\rangle$ is bounded as $\| | \psi_{GS} \rangle - | \psi_D \rangle
\|^2 \le \epsilon_0/L$, with $\epsilon_0$ constant. Now, let $D_L$
be the minimal virtual bond dimension needed for this precision,
i.e. $\| | \psi_{GS} \rangle - | \psi_D \rangle \|^2 \le 2 \times
2L \ \epsilon_{max}(D)$. We demand that
\begin{eqnarray}
\epsilon_{max}(D) & \le & \frac{\epsilon_0}{4 L^2} \nonumber \\
& \le & \exp \left[ \frac{1-\alpha}{\alpha} \left( \frac{c+\tilde
c}{12} \frac{1+\alpha}{\alpha} \log L - \log \frac{D_L}{1-\alpha}
\right) \right],
\end{eqnarray}
from which we can extract
\begin{equation}
D_L \le \mathrm{const} \ \left( \frac{L^2}{\epsilon_0}
\right)^{\frac{\alpha}{1-\alpha}} L^{\frac{c+\tilde c}{12}
\frac{1+\alpha}{\alpha}} \propto {\rm poly} (L).
\end{equation}
$\blacksquare$
\end{proof}

Establishing that there exists an efficient representation of the
ground state is not enough: we must also know if it is possible to
find it efficiently too. In one dimensional gapped systems, the
gap $\Delta$ typically scales polynomially, which means that DMRG
and MPS methods should converge reasonably fast. One can, however,
formalize the regime of efficiency of MPS as a function of how the
different R\'enyi entropies scale with subsystem size
\cite{Schuch07}. In table (\ref{MPStable}) we summarize the
currently known regimes where the MPS approach is an appropriate
one or not.

\begin{table}[t!]
\begin{center}
  \begin{tabular}{| c || c | c | c | c | }
    \hline
    $S_\alpha \sim$ & const & $\log L$ & $L^\kappa (\kappa<1)$ & $L$ \\ \hline
        \hline
    $S_\alpha<1$ & OK & OK & ? & ? \\ \hline
    $S \equiv S_1$ & ? & ? & ? & NO \\ \hline
    $S_\alpha>1$ & ? & ? & NO & NO \\ \hline
  \end{tabular}
\end{center}
   \caption{Relation between scaling of block R\'enyi entropies and approximability by MPS \cite{Schuch07}).
   In the ``undetermined'' region denoted with question marks,
   nothing can be said about approximability just from looking at the scaling.}
   \label{MPStable}
\end{table}

\subsection{Examples}

Here we present a few models and states that are fine examples of
the power of MPS representations\cite{Perez-Garcia07}.

\begin{example}
A well known model with a finite excitation gap and exponentially
decaying spin correlation functions was introduced by Affleck,
Kennedy, Lieb, and Tasaki \cite{Affleck87,Affleck88a}---the so
called AKLT model. The model Hamiltonian is \be H=\sum_i \vec{S}_i
\cdot \vec{S}_{i+1} + \frac{1}{3} \left( \vec{S}_i \cdot
\vec{S}_{i+1} \right)^2. \ee For $S=1$, the local Hilbert space of
each spin has three states, thus $d=3$. The ground state of this
Hamiltonian can be written compactly using a translationally
invariant MPS with bond dimension $D=2$, specifically \be
A_{-1}=\sigma_x, \ A_{0}=\sqrt{2} \sigma^+, \ A_{1}=-\sqrt{2}
\sigma^-. \ee
\end{example}

\begin{example}
A paradigmatic example of a frustrated one dimensional spin chain
is the Majumdar-Ghosh \cite{Majumdar69} model with nearest and next nearest-neighbor interactions: \be
H=\sum_i 2 \vec{\sigma}_i \cdot\vec{\sigma}_{i+1} +
\vec{\sigma}_i\cdot \vec{\sigma}_{i+2}, \ee 
The model is equivalent to the or $J_1-J_2$ Heisenberg
model with $J_1/J_2=2$. The ground state of this model is
composed of singlets between nearest-neighbor spins. However,
since the state must be translationally invariant, we must include
a superposition of singlets between even-odd spins, and
``shifted'' singlets between odd-even spins. The state can be
written compactly in MPS form using $D=3$, \be A_{0}= \left(
\begin{array}{ccc}
0 & 1 & 0  \\
0 & 0 & -1 \\
0 & 0 & 0 \end{array}\right), \ \ A_{1}= \left( \begin{array}{ccc}
0 & 0 & 0  \\
1 & 0 & 0 \\
0 & 1 & 0 \end{array}\right) \ee
\end{example}

\begin{example}
A relevant state for quantum information theory is the
Greenberger--Horne--Zeilinger (GHZ) state, which for $N$ spin $1/2$
particles can be written as \be \ket{GHZ}=\frac{\ket{0}^{\otimes
N}+\ket{1}^{\otimes N}}{\sqrt{2}}. \ee GHZ states are considered
important because for many entanglement measures they are
maximally entangled, however by measuring or tracing out any qubit
a classical state is obtained (although with correlations). GHZ
states can be written using $D=2$ MPS, specifically
$A_{0,1}=1\pm\sigma_z$. Also the ``antiferromagnetic'' GHZ state
is simple, $A_{0,1}=\sigma^\pm$.
\end{example}

\begin{example}
Cluster states are relevant for one-way quantum computing. They
are the ground state of \be H=\sum \sigma^z_{i-1} \sigma^x_i
\sigma^z_{i+1}, \ee and can be represented using a $D=2$ MPS, \be
A_{0}= \left( \begin{array}{cc}
0 & 0  \\
1 & 1 \end{array}\right), \ \ A_{1}= \left( \begin{array}{cc}
1 & -1 \\
0 & 0 \end{array}\right) \ee
\end{example}

\begin{example}
(Classical superposition MPS) Imagine we have a classical
Hamiltonian \be H=\sum_{(i,j)} h(\sigma_i, \sigma_j), \ee where
$\sigma_i=1,\ldots,d$, and $h(\sigma_i, \sigma_j)$ are local
interactions. The partition function of such a model at a given
inverse temperature $\beta$ is \be Z=\sum_{\{\sigma\}} \exp
\left[-\beta H(\sigma) \right], \ee where the sum is over all
possible configurations of the vector $\sigma$. Let us now define
a quantum state $\ket{\psi_\beta}$ whose amplitude for a given
state of the computational basis corresponds to the term in the
partition function for that state, i.e.
\begin{eqnarray}
\ket{\psi_\beta}&=&\frac{1}{\sqrt{Z}} \sum_{\{\sigma\}} \exp \left[-\frac{\beta}{2} H(\sigma) \right] \ket{\sigma_1\ldots\sigma_N} \nonumber \\
&=&\frac{1}{\sqrt{Z}} \sum_{\{\sigma\}} \prod_{(i,j)} \exp
\left[-\frac{\beta}{2} h(\sigma_i,\sigma_j) \right]
\ket{\sigma_1\ldots\sigma_N}.
\end{eqnarray}
We shall now define a map $P$ ---in the same manner as in valence
bond states--- that goes from $\mathbb{C}^{d^2}$ to $\mathbb{C}^2$
such that \be P\ket{s,k}=\ket{s} \langle \varphi_s | k\rangle, \ee
where we have defined \be \sum_{\alpha=1}^d \langle \varphi_s |
\alpha \rangle \langle \varphi_{\tilde s} | \alpha \rangle = \exp
\left[-\frac{\beta}{2} h(s,{\tilde s}) \right]. \ee To visualize
what happens when we insert these back into the classical
superposition state $\ket{\psi_\beta}$, let us concentrate for a
moment on a one-dimensional system:
\begin{eqnarray}
\ket{\psi_\beta}&=&\frac{1}{\sqrt{Z}}
\sum_{\sigma_1,\ldots,\sigma_N} \exp \left[-\frac{\beta}{2}
h(\sigma_1,\sigma_2) \right]
\ldots\exp \left[-\frac{\beta}{2} h(\sigma_{N-1},\sigma_N) \right] \ket{\sigma_1\ldots\sigma_N}, \nonumber \\
\ket{\psi_\beta}&=&\frac{1}{\sqrt{Z}}
\sum_{\sigma_1,\ldots,\sigma_N}\left( \sum_{\alpha_1=1}^d \langle
\varphi_{\sigma_1} |  \alpha_1 \rangle \langle \varphi_{\sigma_2}
| \alpha_1 \rangle
\sum_{\alpha_2=1}^d \langle \varphi_{\sigma_2} | \alpha_2\rangle \langle \varphi_{\sigma_3} | \alpha_2\rangle\right.\nonumber\\
&&\hspace{2.3cm}\left.\times\ldots\times\sum_{\alpha_N=1}^d
\langle \varphi_{\sigma_N} | \alpha_N\rangle
\langle \varphi_{\sigma_1} | \alpha_N\rangle\right)\ket{\sigma_1\ldots\sigma_N}, \\
\ket{\psi_\beta}&=&\frac{1}{\sqrt{Z}}
\sum_{\sigma_1,\ldots,\sigma_N}
\sum_{\alpha_1,\ldots,\alpha_N=1}^{d}
%
%
\left[\langle \varphi_{\sigma_1} |  \alpha_N \rangle\langle
\varphi_{\sigma_1} |  \alpha_1 \rangle\right]
\left[ \langle \varphi_{\sigma_2} | \alpha_1 \rangle \langle
\varphi_{\sigma_2} | \alpha_2\rangle \right]\nonumber\\
&&\hspace{3.45cm}\times\ldots\times \left[ \langle
\varphi_{\sigma_N} | \alpha_{N-1}\rangle \langle
\varphi_{\sigma_N} | \alpha_N\rangle\right]
\ket{\sigma_1\ldots\sigma_N},
\end{eqnarray}
and we can replace $A^{(i)}_{s_i,\alpha,\beta} = \langle
\varphi_{s_i} | \alpha \rangle \langle \varphi_{s_i} | \beta
\rangle$, thus expressing the classical thermal superposition
state as a MPS.

\renewcommand{\labelenumi}{(\roman{enumi})}

These states have some important properties:
\begin{enumerate}
\item They obey strict area laws,
\item They allow to calculate classical and quantum correlations, and
\item They are ground states of local Hamiltonians.
\end{enumerate}

Property (i) should be obvious by now, since we have explicitly
shown the MPS form of the state. We can show easily property (ii)
for Ising models. A classical correlation function $f$ must be
evaluated with the partition function, $\langle
f(\sigma)\rangle=\sum_{\sigma} f(\sigma) e^{-\beta H(\sigma)}/Z$,
but this is just the expectation value of an operator made of
changing the argument of $f$ into $\sigma_z$ operators, and
evaluated with $\ket{\psi_\beta}$. Since it is the expectation
value of a MPS, it is efficient to compute. Finally, we will
demonstrate property (iii) at length in the next section, because
it will lead us into the final topic of this lectures: Quantum
kinetic models.
\end{example}

\subsection{Classical kinetic models}

Our goal in this section is to show the local Hamiltonians whose
ground state is the classical superposition state defined in the
previous section. As we shall see, these Hamiltonians will arise
from the master equation of a classical system that is interesting
in its own right, so we will first spend some time on it.

Let us consider a system made out of $N$ classical spins
interacting through a Hamiltonian $H$. If  $\sigma_{i}$ denotes
the state of $i$th spin, we will label the configurations of the
system by $\sigma=(\sigma_{1},\ldots,\sigma_{N})$, and the
probability of finding at time $t$ the system in state $\sigma$
(given that it was in state $\sigma_0$ at time $t_0$) by
$P(\sigma,t)=P(\sigma,t|\sigma_0,t_0)$. In what follows we focus
on this probability distribution, whose dynamics is described by a
master equation:
\be\label{masterEq} {\dot P}(\sigma,t)=\sum_{\sigma' }
W(\sigma,\sigma') P(\sigma',t)
   - \sum_{\sigma'} W(\sigma',\sigma) P(\sigma,t),
\ee
where $W(\sigma,\sigma')$ is the transition probability from state
$\sigma'$ to state $\sigma$. This equation defines the class of
kinetic models, and it clearly describes a Markov process --- the
instantaneous change of $P(\sigma,t)$ does not depend on its
history.

We will only consider systems that obey a detailed balance
condition, i.e.
\be\label{detailed} W(\sigma,\sigma') e^{-\beta H(\sigma')} =
W(\sigma',\sigma) e^{-\beta H(\sigma)}. \ee
With this condition, the stationary state of the master equation
(the one that fulfills ${\dot P}_{\mathrm{st}}(\sigma,t)=0$) is
simply $P_{\mathrm{st}}(\sigma)=e^{-\beta H(\sigma)}/Z$, with $Z$
being the partition function. This state in particular will map
into the classical superposition state defined above, but we still
have not found its parent Hamiltonian. For this, we will rewrite
Eq.~(\ref{masterEq}) in the form of a matrix Schr\"odinger
equation (albeit with imaginary time) from which we can identify a
Hamiltonian.

Let us apply the transformation $\psi(\sigma,t)=e^{\beta
H(\sigma)/2}P(\sigma,t)$, which leads to
\begin{eqnarray}
{\dot \psi}(\sigma,t)&=&
   \sum_{\sigma'} e^{\beta H(\sigma)/2}W(\sigma,\sigma') e^{-\beta H(\sigma')/2} \psi(\sigma',t)
   -W(\sigma',\sigma) \psi(\sigma,t)\nonumber\\
&=&-\sum_{\sigma'}H_\beta(\sigma,\sigma')\psi(\sigma',t)
\end{eqnarray}
with
\begin{equation}
H_\beta(\sigma,\sigma')=\sum_{\sigma''}W(\sigma'',\sigma')\delta_{\sigma\sigma'}-e^{\beta
H(\sigma)/2}W(\sigma,\sigma')e^{-\beta H(\sigma')/2}.
\end{equation}
Notice that the detailed balance condition guarantees that the
matrix $H_{\beta}$ is Hermitian, so we can interpret it as a
Hamiltonian. Furthermore, because of the conservation of
probability, $H_\beta$ can only have non-negative eigenvalues,
which means that the state $\psi_{\mathrm{st}}$ associated to the
stationary state $P_{\mathrm{st}}$ with eigenvalue zero must be a
ground state: the classical superposition MPS that we were looking
for.

Remarkably, we have not said anything yet about $H$. A famous
example of such kinetic model is a single spin-flip model
considered by Glauber \cite{Glauber63JMP}, for which $H$ is the
Ising Hamiltonian\footnote{This is the reason why the Glauber
model is also known as the kinetic Ising model (KIM).}
\begin{equation}
H(\sigma)\equiv H_{\mathrm{Ising}}(\sigma)=-J\sum_{\langle
i,j\rangle}\sigma_{i}^{z}\sigma_{j}^{z}\qquad (J>0).
\end{equation}
Denoting by $P_{i}$ the flip operator of the $i$-th spin, i.e.,
$P_i \sigma_i=-\sigma_i$, the general master equation
(\ref{masterEq}) reduces in this case to
\begin{equation}\label{SingleFlip}
\dot{P}(\sigma,t)=\sum_{i}\left[W(\sigma,P_{i}\sigma)P(P_{i}\sigma,t)-W(P_{i}\sigma,\sigma)P(\sigma,t)\right]
\end{equation}
with $W(\sigma,P_{i}\sigma)$ now called spin rates. It was shown
in \cite{Glauber63JMP} that the most general form of spin rates
with symmetric interaction with both nearest-neighbors, and
satisfying the detailed balance condition (\ref{detailed}), is
given by
\begin{equation}\label{SpinRates1}
w(P_{i}\sigma,\sigma)=\Gamma(1+\delta\sigma_{i-1}\sigma_{i+1})[1-(1/2)\gamma\sigma_{i}(\sigma_{i-1}+\sigma_{i+1})]
\end{equation}
with $\Gamma>0$, $-1\leq \delta\leq 1$, and $0\leq \gamma\leq 1$.
The $\delta=0$ case was thoroughly investigated by Glauber
\cite{Glauber63JMP}, who showed that all the relevant quantities
can be derived analytically --- including the dynamical exponent
that turned out to be $z=2$. The more general case of nonzero
$\delta$ was treated in a series of papers
\cite{DekerHaake79,Kimball79,HaakeThol80}, that showed for
instance that the choice $\delta=\gamma/(2-\gamma)$ leads to an
interesting dynamical exponent $z\neq 2$.

If we rewrite the single spin-flip master equation in the form of
the Schr\"odinger equation, we obtain an associated quantum
Hamiltonian
\begin{eqnarray}\label{HamilFelderhof}
H_\beta(\delta,\gamma)&=&-\Gamma\sum_{i}\left[\left(A(\delta,\gamma)-B(\delta,\gamma)\sigma_{i-1}^{z}\sigma_{i+1}^{z}\right)\sigma_{i}^{x}\right.\nonumber\\
&&\left.\hspace{1.4cm}-(1+\delta\sigma_{i-1}^{z}\sigma_{i+1}^{z})\left(1-(1/2)\gamma\sigma_{i}^{z}\left(\sigma_{i-1}^{z}+\sigma_{i+1}^{z}\right)\right)\right],\nonumber\\
\end{eqnarray}
where
\begin{equation}
A(\delta,\gamma)=\frac{(1+\delta)\gamma^{2}}{2(1-\sqrt{1-\gamma^{2}})}-\delta,
\qquad B(\delta,\gamma)=1-A(\delta,\gamma)
\end{equation}
and $\sigma^{z}$ and $\sigma^{x}$ are the standard Pauli matrices.
For $\delta=0$ this Hamiltonian was diagonalized in Ref.
\cite{Felderhof71}, and independently in Ref.\ \cite{Siggia77}.

The Hamiltonian $H_\beta(\delta,\gamma)$, and also the other ones
that can be derived in this way, are typically gapped except at a
critical temperature $\beta_c$ where the gap vanishes with the
critical exponent $z$ that characterizes the model. In one
dimension $\beta_c=\infty$, but for larger dimensions this model
has a finite critical temperature.

We have seen thus far how the the master equation of a classical
spin model (that obeys the detailed balance condition) can be
associated to a quantum Hamiltonian with some interesting critical
properties --- for example, its ground state obeys a strict area
law and can be written efficiently as a MPS. Nevertheless, the
underlying model is still classical. In the next section, we will
see one way in which we can generalize the initial model to be
quantum, while retaining the same structure that leads to
associated Hamiltonians that obey area laws.

\section{Quantum kinetic Ising models}

Here we discuss ways to generalize the kinetic equation
(\ref{masterEq}) to a quantum master equation, but in such a way
that its diagonal part reproduces the corresponding kinetic model.
A similar approach was taken in Ref.\ \cite{Heims}, where a
quantum master equation that reproduced a kinetic Ising model was
proposed (see also Ref.\ \cite{Kawasaki}). However, no attempts
aiming at fully solving such QMEs are known so far. Our purpose is
to give quantum generalizations of the classical kinetic models
that can be solved analytically.

Recently, we presented such a generalization \cite{Augusiak10} for
the single spin-flip model, Eq.\ (\ref{SingleFlip}), with the spin
rates of Eq.\ (\ref{SpinRates1}). In Ref.\ \cite{Augusiak10} we were
able to decouple the master equation for the density matrix of a
quantum system into $2^N$ master equations with the same structure
as the ones studied above. Here, we will only show the associated
Hamiltonians (and their spectra) obtained in these models.
However, we will demonstrate how to approach the problem but
in a different model that allows transitions that flip
two consecutive spins.

\subsection{A two spin flip model}

First, let us particularize the classical kinetic equation
(\ref{masterEq}) to the case where the flip operator acts on pairs
of consecutive spins of the chain, i.e.
\begin{equation}
\frac{\partial P(\sigma,t)}{\partial
t}=\sum_{i}\left[w_{i}(F_{i,i+1}\sigma\rightarrow
\sigma)P(F_{i,i+1}\sigma,t)-w_{i}(\sigma\rightarrow
F_{i,i+1}\sigma)P(\sigma,t)\right],
\end{equation}
where $F_{i,i+1}$ denote spin flips at positions $i$ and $i+1$,
while the spin rates are given by
$w_{i}(F_{i,i+1}\sigma,\sigma)=\Gamma[1-(1/2)\gamma(\sigma_{i-1}\sigma_{i}
+\sigma_{i+1}\sigma_{i+2})]$ with $0<\Gamma<\infty$ and
$\gamma=\tanh2\beta J$. This model was investigated in Ref.
\cite{HilhorstSuzukiFelderhof}, where the associated Hamiltonian
was found and diagonalized using the Jordan-Wigner transformation
\cite{JordanWigner} followed by Fourier and Bogoliubov-Valatin
\cite{Bogoliubov,Valatin} transformations. In particular, Hilhorst
{\it et al.} were able to show that, despite the complexity of the
transformations, one can easily compute expectation values such as
magnetization, energy density, or correlations, and that they have
a relatively simple exponential behavior
\cite{HilhorstSuzukiFelderhof}.

Here we will define through a master equation a quantum model that
resembles the kinetic model above. For this, we will replace
classical probabilities with the quantum density matrices, and
classical operators with quantum ones (e.g. $\sigma^x$ is the
qubit flip operator). Consider the following master equation
\begin{equation} \label{Qmaster}
\partial_{t}\varrho(t)=\sum_{i}\left[\sigma_{i}^{x}\sigma_{i+1}^{x}
\sqrt{w_{i}(\sigma^{z})}\varrho(t)\sqrt{w_{i}(\sigma^{z})}\,\sigma_{i}^{x}\sigma_{i+1}^{x}
-\frac{1}{2}\{w_{i}(\sigma^{z}),\varrho(t)\}\right],
\end{equation}
where $\{\cdot,\cdot\}$ denotes the anticommutator and
$w_{i}(\sigma^{z})$ are quantum mechanical generalizations of the
spin rates (\ref{SpinRates1}), now written in terms of the
$\sigma^z$ operators,
\begin{equation}
w_{i}(\sigma^{z})=\Gamma\left[1-\textstyle\frac{1}{2}\gamma(\sigma_{i-1}^{z}\sigma_{i}^{z}
+\sigma_{i+1}^{z}\sigma_{i+2}^{z})\right].
\end{equation}

Although it looks complicated, the quantum kinetic model above can
still be solved with techniques similar to the classical
case \cite{Augusiak10}: the key ingredient is to find a large
number of constants of motion that allow to split the master
equation into a set of ordinary Schr\"odinger equations. To see
this, we must represent the density matrix $\varrho(t)$ as a
vector in an expanded Hilbert space. This follows from a simple
isomorphism between linear operators from $M_{d}(\mathbb{C})$ and
vectors from $\mathbb{C}^{d^{2}}$. In other words, writing our
density matrix in the computational basis in
$(\mathbb{C}^{2})^{\ot N}$ as
$\varrho(t)=\sum_{\sigma,\widetilde{\sigma}}[\varrho(t)]_{\sigma,\widetilde{\sigma}}\ket{\sigma}\!\bra{\widetilde{\sigma}}$,
we can treat it as a vector
$\ket{\varrho(t)}=\sum_{\sigma,\widetilde{\sigma}}[\varrho(t)]_{\sigma,\widetilde{\sigma}}\ket{\sigma}\ket{\widetilde{\sigma}}$
from $(\mathbb{C}^{2})^{\ot N}\ot (\mathbb{C}^{2})^{\ot N}$. Even
if formally we are enlarging the number of spins from $N$ to $2N$,
the advantage is that now we deal with ``pure states" instead of
density matrices which allows us to find many conserved
quantities. This, in turn, shows that the effective Hilbert space
used is much smaller than the initial one.
To be consistent, operators that appear to the right of
$\varrho(t)$ must be replaced with ``tilded'' operators that act
on the right subsystem of the expanded space, while operators on
the left of the density matrix (``untilded'') act on the left
subsystem (for instance
$\sigma_{i}^{x}\widetilde{\sigma}_{i}^{x}\ket{s}\ket{\widetilde{s}}=
\sigma_{i}\ket{s}\widetilde{\sigma}_{i}^{x}\ket{\widetilde{s}}$).
This notation allows us to rewrite the master equation
(\ref{Qmaster}) as the following matrix equation
\begin{equation}\label{vector_eq}
\ket{\dot{\varrho}(t)}=
\sum_{i}\left[\sigma_{i}^{x}\sigma_{i+1}^{x}\widetilde{\sigma}_{i}^{x}\widetilde{\sigma}_{i+1}^{x}\sqrt{w_{i}(\sigma^{z})w_{i}(\widetilde{\sigma}^{z})}
-\frac{1}{2}[w_{i}(\sigma^{z})+w_{i}(\widetilde{\sigma}^{z})]\right]\ket{\varrho(t)}.
\end{equation}
As was the case for the initial classical master equation, the
matrix appearing on the right-hand side of Eq.\ (\ref{vector_eq})
is not Hermitian. In order to bring it to Hermitian form we can
use the detailed balance condition, which suggests the
transformation
\begin{equation}\label{transformation}
\ket{\varrho(t)}= \exp\left[-(\beta/4)[\mathcal{H}(\sigma)+
\mathcal{H}(\widetilde{\sigma})]\right]\ket{\psi(t)},
\end{equation}
with $\mathcal{H}$ denoting the quantum generalization of the
Ising Hamiltonian
$\mathcal{H}=-J\sum_{i}\sigma_{i}^{z}\sigma_{i+1}^{z}$. With this
transformation, and denoting
\begin{equation}
v_{i}(\sigma^{z})=w_{i}(\sigma^{z})\exp[(\beta
J)\sigma_{i}^{z}(\sigma_{i-1}^{z}+\sigma_{i+1}^{z})],
\end{equation}
Eq.\ (\ref{vector_eq}) can be written as
\begin{equation}
\label{twoflipSchrodinger} \ket{\dot{\psi}(t)}=
\sum_{i}\left[\sigma_{i}^{x}\sigma_{i+1}^{x}\widetilde{\sigma}_{i}^{x}\widetilde{\sigma}_{i+1}^{x}\sqrt{v_{i}(\sigma^{z})v_{i}(\widetilde{\sigma}^{z})}
-\frac{1}{2}[v_{i}(\sigma^{z})+v_{i}(\widetilde{\sigma}^{z})]\right]\ket{\psi(t)}
\end{equation}
which we can see as a Schr\"odinger equation
$\ket{\dot{\psi}(t)}=-H\ket{\psi(t)}$ with Hermitian $H$.

We have reached the point where all these changes of notation
payoff: indeed, the form of $H$ makes it clear that it commutes
with
$\sigma_{i}^{z}\sigma_{i+1}^{z}\widetilde{\sigma}_{i}^{z}\widetilde{\sigma}_{i+1}^{z}$
$(i=1,\ldots,N)$. Therefore, we can introduce new variables
$\tau_{i}=\sigma_{i}^{z}\sigma_{i+1}^{z}\widetilde{\sigma}_{i}^{z}\widetilde{\sigma}_{i+1}^{z}$
$(i=1,\ldots,N)$ which are constants of motion and reduce the
number of degrees of freedom. In particular, tilded variables can
be expressed by  $\sigma$  and the new variables $\tau$ as
$\widetilde{\sigma}_{i}^{z}\widetilde{\sigma}_{i+1}^{z}=\tau_{i}\sigma_{i}^{z}\sigma_{i+1}^{z}$
for any $i$. In other words, we have replaced $\sigma$ and
$\widetilde{\sigma}$ by $\tau$ and $\sigma$, of which $\tau$ is
conserved. To each configuration of $\tau$'s we associate a
natural number from 0 to $2^N-1$, which corresponds to a
particular correlation between the $\sigma$ and
$\widetilde{\sigma}$ variables. For example, $\tau=0$ corresponds
to all $\tau$-spins up ($\tau_{i}=1$ for $i=1,\ldots,N$), while
$\tau=2^{N}-1$ means that $\tau_{i}=-1$ for $i=1,\ldots,N$. With
this notation, each value of $\tau$ is associated to a Hamiltonian
$H_{\tau}$ that acts only in the space of $N$ spins and is of the
form
\begin{eqnarray}\label{Htau}
H_{\tau}&=&-\sum_{i}
\left[\sigma_{i}^{x}[v_{i}(\sigma^{z})]^{\frac{1}{2}}[v_{i}(\tau\sigma^{z})]^{\frac{1}{2}}
-\frac{1}{2}[w_{i}(\sigma^{z})+w_{i}(\tau\sigma^{z})]\right],
\end{eqnarray}
where $\tau\sigma^{z}$ denotes $\tau_{i}\sigma_{i}^{z}$
$(i=1,\ldots,N)$. Because these Hamiltonians are independent from
each other, we have converted the problem of solving the general
master equation (\ref{Qmaster}) to the problem of diagonalizing
$2^{N}$ Hamiltonians, each of dimension $2^{N}\times 2^{N}$. Now,
we have that
\begin{equation}
\ket{\psi(t)}=\bigotimes_{\tau=0}^{2^{N}-1}\ket{\psi_{\tau}(t)},\qquad
H=\bigotimes_{\tau=0}^{2^{N}-1}H_{\tau}.
\end{equation}
After simple algebra one sees that the explicit form of $H_{\tau}$
is
\begin{eqnarray}\label{HamilTwo}
H_{\tau}&=&-\sum_{i}\Big[\left(A_{i}(\varphi)-B_{i}(\varphi)\sigma_{i-1}^{z}
\sigma_{i}^{z}\sigma_{i+1}^{z}\sigma_{i+2}^{z}\right)
\sigma_{i}^{x}\sigma_{i+1}^{x}\nonumber\\
&&\hspace{1cm}\left.-\left[1-\frac{1}{2}\gamma\left(f(\tau_{i-1})\sigma_{i-1}^{z}
\sigma_{i}^{z}+f(\tau_{i+1})\sigma_{i+1}^{z}\sigma_{i+2}^{z}\right)\right]\right],
\end{eqnarray}
where
\begin{equation}\label{ABTwo}
A_{i}(\varphi)=\left\{
\begin{array}{ll}
\cos^{2}\!\varphi, &\;\; \tau_{i-1}\tau_{i+1}=1\\[1ex]
\sqrt{\cos 2\varphi}, &\;\; \tau_{i-1}\tau_{i+1}=-1
\end{array}
\right.,\qquad B_{i}(\varphi)=\left\{
\begin{array}{ll}
\sin^{2}\!\varphi, &\;\; \tau_{i-1}\tau_{i+1}=1\\[1ex]
0, &\;\; \tau_{i-1}\tau_{i+1}=-1,
\end{array}
\right.
\end{equation}
with
\begin{equation}
\cos\varphi=\frac{\cosh\!\beta J}{(\cosh^{2}\!\beta
J+\sinh^{2}\!\beta J)^{1/2}},\qquad \sin\varphi=\frac{\sinh\!\beta
J}{(\cosh^{2}\!\beta J+\sinh^{2}\!\beta J)^{1/2}},
\end{equation}
and  $f(x)=(1/2)(1+x)$. Here the angle ranges from zero (which
corresponds to infinite temperature) to $\pi/4$ (which corresponds
to $T=0$) and in this notation $\gamma=\sin2\varphi$. Let us
notice that for $\tau=0$ Eqs. (\ref{HamilTwo}) and (\ref{ABTwo}),
as it should be, reproduce the Hamiltonian derived in
\cite{HilhorstSuzukiFelderhof}. This, however, contrary to the
single spin-flip case, is not the case for $\tau=2^{N}-1$, where
one of the terms in the square brackets vanishes and the
Hamiltonian reduces to
\begin{equation}
H_{2^{N}-1}=-\sum_{i}\left[\left(A_{i}(\varphi)-B_{i}(\varphi)\sigma_{i-1}^{z}
\sigma_{i}^{z}\sigma_{i+1}^{z}\sigma_{i+2}^{z}\right)
\sigma_{i}^{x}\sigma_{i+1}^{x}-1\right]
\end{equation}

Let us discuss now some of the properties of $H_{\tau}$. Below we
show that for all $\tau$ they are always positive operators. We
also find all the cases with respect to $\varphi$ and $\tau$ for
which the Hamiltonians can have zero-energy ground states.

\begin{lemma}\label{LemmaHamil}
The Hamiltonians $H_{\tau}$ are positive for any
$\tau=0,\ldots,2^{N}-1$.
\end{lemma}
\begin{proof}
Let us denote by $H_{\tau}^{(i)}$ the $i$th term appearing in the
sum in Eq.\ (\ref{HamilTwo}). The idea is to show that all
$H_{\tau}^{(i)}$ are positive, and the positivity of $H_{\tau}$
follows immediately. Of course, the form of $H_{\tau}^{(i)}$
changes depending on $\tau$-spins at positions $i-1$ and $i+1$.
Therefore, we distinguish several cases with respect to different
possible configurations of these spins.

For $\tau_{i-1}=\tau_{i+1}$, one easily infers from Eqs.
(\ref{HamilTwo}) and (\ref{ABTwo}) that
\begin{eqnarray}
H_{\tau}^{(i)}&=&1-\frac{1}{2}\gamma\left[f(\tau_{i-1})\sigma_{i-1}^{z}\sigma_{i}^{z}
+f(\tau_{i+1})\sigma_{i+1}^{z}\sigma_{i+2}^{z}\right]\nonumber\\
&&-\left(\cos^{2}\!\varphi-\sin^{2}\!\varphi\,\sigma_{i-1}^{z}\sigma_{i}^{z}
\sigma_{i+1}^{z}\sigma_{i+2}^{z}\right)\sigma_{i}^{x}\sigma_{i+1}^{x}.
\end{eqnarray}
In the case when both spins $\tau_{i-1}$ and $\tau_{i+1}$ are
down, the function $f$ is zero and both terms in square brackets
vanish and the above operator becomes
$1-(\cos^{2}\varphi-\sin^{2}\sigma_{i-1}^{z}\sigma_{i}^{z}
\sigma_{i+1}^{z}\sigma_{i+2}^{z})\sigma_{i}^{x}\sigma_{i+1}^{x}$.
It is clear then that its minimal eigenvalue is zero. In the case
when $\tau_{i-1}=\tau_{i+1}=1$ these terms do not vanish, however,
still this is effectively a $16\times 16$ matrix which can be
shown to be positive computationally: using the software
Mathematica we can easily see that the minimal eigenvalue is zero.

For $\tau_{i-1}=-\tau_{i+1}$, one of the values $f(\tau_{i-1})$ or
$f(\tau_{i+1})$ is zero. Assuming that $f(\tau_{i-1})=0$ (the case
of $f(\tau_{i+1})=0$ leads to the same eigenvalues), one has
\begin{equation}
H_{\tau}^{(i)}=1-\frac{1}{2}\sin2\varphi\,\sigma_{i+1}^{z}\sigma_{i+2}^{z}
-\sqrt{\cos2\varphi}\,\sigma_{i}^{x}\sigma_{i+1}^{x}.
\end{equation}
When constrained to three consecutive spins ($i-1$, $i$, and
$i+1$) this $H_{\tau}^{(i)}$ is just a $8$ by $8$ matrix (on the
remaining spins it acts as the identity matrix) and its
eigenvalues can be obtained using Mathematica. One then checks
that its minimal eigenvalue is
$1-(1/2)\sqrt{4\cos2\varphi+(\sin2\varphi)^{2}}$ with
$\varphi\in[0,\pi/4]$. Simple analysis shows that this is a
nonnegative function of $\varphi$ and gives zero only when
$\varphi=0$.

In conclusion, $H_{\tau}^{(i)}\geq 0$ for all $\tau$s and
$\varphi\in[0,\pi/4]$ and therefore our Hamiltonians $H_{\tau}$
are positive. $\blacksquare$
\end{proof}

\begin{figure}[t!]
\begin{center}
\includegraphics[width=0.70\textwidth]{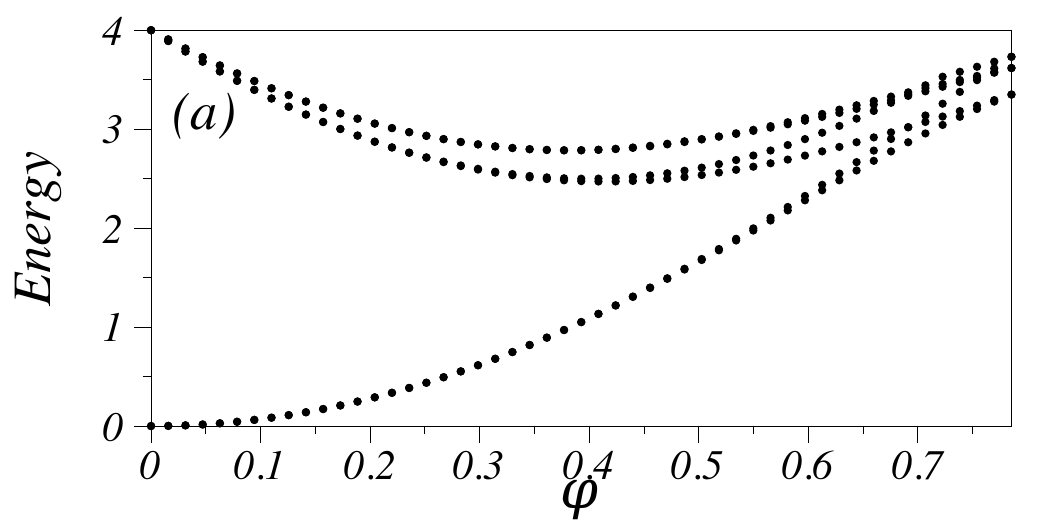}
\includegraphics[width=0.70\textwidth]{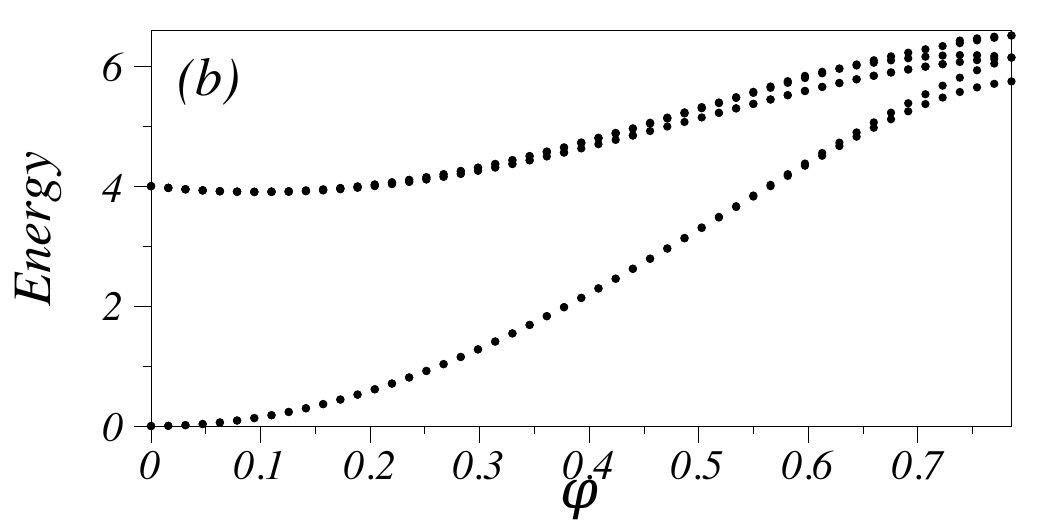}
\includegraphics[width=0.70\textwidth]{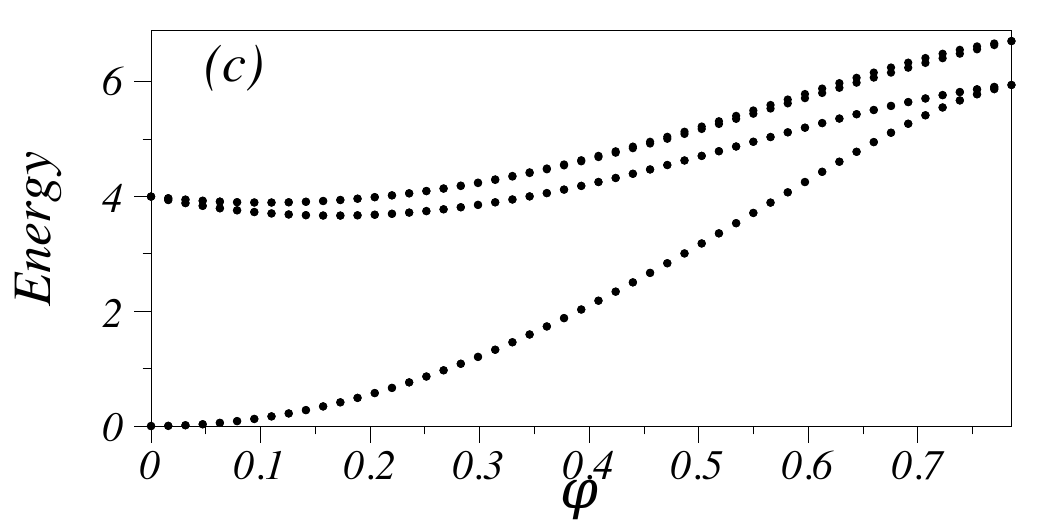}
\caption{Low energy states of the Hamiltonians (\ref{HamilTwo})
associated to the two flip quantum master equation for a system
with $N=16$ spins as a function of $\varphi$. The panels are (a)
$\tau=2^8-1$ (half $\tau$-spins up and half down), (b) $\tau=2^8$
(only one $\tau$-spin up, the others down), and (c) $\tau=2^9+2^8$
(two neighboring $\tau$-spins up, the others down). Only in case
(c) the ground state is fully degenerate for all values of
$\varphi$, in the other two the first excited state energy is very
close but not equal to the ground state. In case (b) the ground
state is not degenerate at $\varphi=\pi/4$, while in the other two
cases it is. } \label{fig:spectra}
\end{center}
\end{figure}

Based on the above analysis, let us now distinguish all the cases
with respect to $\tau$ and $\varphi$ when $H_{\tau}(\varphi)$ can
have zero-energy eigenstates. It clearly follows from the proof of
lemma \ref{LemmaHamil} that if $\tau\neq 0$ or $\tau\neq 2^{N}-1$
there exists $i$ such that $\tau_{i-1}\neq \tau_{i+1}$ and then
the corresponding $H_{\tau}(\varphi)$ can have zero eigenvalues
only when $\varphi=0$. Let us now discuss this case. It follows
from Eqs. (\ref{HamilTwo}) and (\ref{ABTwo}) that for $\varphi=0$,
which corresponds to infinite temperature, the dependence on
$\tau$ vanishes and one obtains
\begin{equation}
H_{\tau}(0)\equiv
\overline{H}=\sum_{i}\left(1-\sigma_{i}^{x}\sigma_{i+1}^{x}\right),
\end{equation}
which has a doubly degenerate ferromagnetic ground state.

For $\tau=0$ one gets the Hamiltonian obtained in
\cite{HilhorstSuzukiFelderhof}, that is
\begin{eqnarray}
H_{0}(\varphi)&=&-\sum_{i}
\left[\left(\cos^{2}\!\varphi-\sin^{2}\!\varphi\,\sigma_{i-1}^{z}\sigma_{i}^{z}\sigma_{i+1}^{z}\sigma_{i+2}^{z}\right)
\sigma_{i}^{x}\sigma_{i+1}^{x}\right.\nonumber\\
&&\hspace{1cm}-\left.\left(1-(1/2)\sin2\varphi(\sigma_{i-1}^{z}\sigma_{i}^{z}+\sigma_{i+1}^{z}\sigma_{i+2}^{z})\right)\right].
\end{eqnarray}
The ground state of this Hamiltonian is  doubly degenerate for all
values of $\varphi$, except for $\varphi=\pi/4$ (zero temperature)
where also the first excited state becomes degenerate with the
ground state \cite{HilhorstSuzukiFelderhof}. For many values of
$\tau$ this statement holds, except that the ground state has a
positive energy --- implying that the off-diagonal elements of the
QME decay in time. In other cases we find that the ground state is
unique for all values of $\varphi$, even $\pi/4$. Typical spectra
for some values of $\tau$ in finite systems are shown in Fig.\ \ref{fig:spectra}.

\subsection{The single flip model}

For comparison only, we reproduce here the associated Hamiltonians
that are obtained when single flip processes are allowed in the
quantum master equation \cite{Augusiak10}.
Again, a set of conserved quantities allows us to break the QME
into $2^N$ Schr\"odinger equations labeled by a parameter $\tau$
 \begin{equation}\label{SchrEq}
 \ket{\psi_{\tau}(t)}=-H_{\tau}\ket{\psi_{\tau}(t)}\qquad
 (\tau=0,\ldots,2^{N}-1),
 \end{equation}
 where the Hamiltonians $H_{\tau}$ are given by
 \begin{eqnarray}\label{singleHamil}
 H_{\tau}\equiv
 H_{\tau}(\delta,\gamma)&=&-\Gamma\sum_{i}\left[\left(\widetilde{A}_{i}(\delta,\gamma)-\widetilde{B}_{i}(\delta,\gamma)
 \sigma_{i-1}^{z}\sigma_{i+1}^{z}\right)\sigma_{i}^{x}\right.\nonumber\\
 &&\hspace{1.3cm}-1+\frac{\gamma}{2}(1+\delta)\sigma_{i}^{z}
 \left(f(\tau_{i-1}\tau_{i})\sigma_{i-1}^{z}+f(\tau_{i}\tau_{i+1})\sigma_{i+1}^{z}\right)\nonumber\\
 &&\left.\hspace{1.25cm}-\delta
 f(\tau_{i-1}\tau_{i+1})\sigma_{i-1}^{z}\sigma_{i+1}^{z}\right],
 \end{eqnarray}
 where
 \begin{eqnarray}\label{Ai}
 \widetilde{A}_{i}(\gamma,\delta)=\left\{
 \begin{array}{ll}
 \displaystyle\frac{(1+\delta)\gamma^{2}}{2(1-\sqrt{1-\gamma^{2}})}-\delta, & \quad\tau_{i-1}=\tau_{i+1},\\[3ex]
 \sqrt{1-\delta^{2}}\sqrt[4]{1-\gamma^{2}}, &\quad
 \tau_{i-1}=-\tau_{i+1}
 \end{array}
 \right.
 \end{eqnarray}
 and
 \begin{eqnarray}\label{Bi}
 \widetilde{B}_{i}(\gamma,\delta)=\left\{
 \begin{array}{ll}
 1-\displaystyle\frac{(1+\delta)\gamma^{2}}{2(1-\sqrt{1-\gamma^{2}})}, &\quad \tau_{i-1}=\tau_{i+1},\\[3ex]
 0, &\quad \tau_{i-1}=-\tau_{i+1}.
 \end{array}
 \right.
 \end{eqnarray}
 Here, each $\tau$ means a configuration of the conserved quantities
 that is different than the one shown above for the two flip model --
 however, we still use its binary representation so that $\tau$
  is a shorthand notation for $N$ variables
 $(\tau_{1},\ldots,\tau_{N})$, each taking values $\pm 1$.
 The equation in (\ref{SchrEq}) for $\tau=0$
 corresponds to the diagonal elements of $\varrho(t)$, while for
 the remaining $\tau\neq 0$, they describe the off-diagonal
 elements of the density matrix.

\begin{figure}[t!]
\begin{center}
\includegraphics[width=0.70\textwidth]{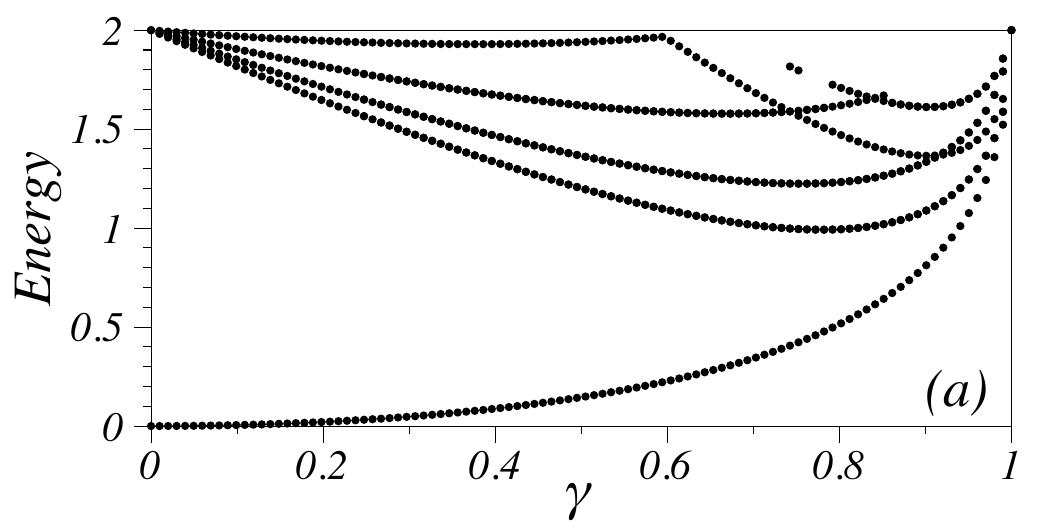}
\includegraphics[width=0.70\textwidth]{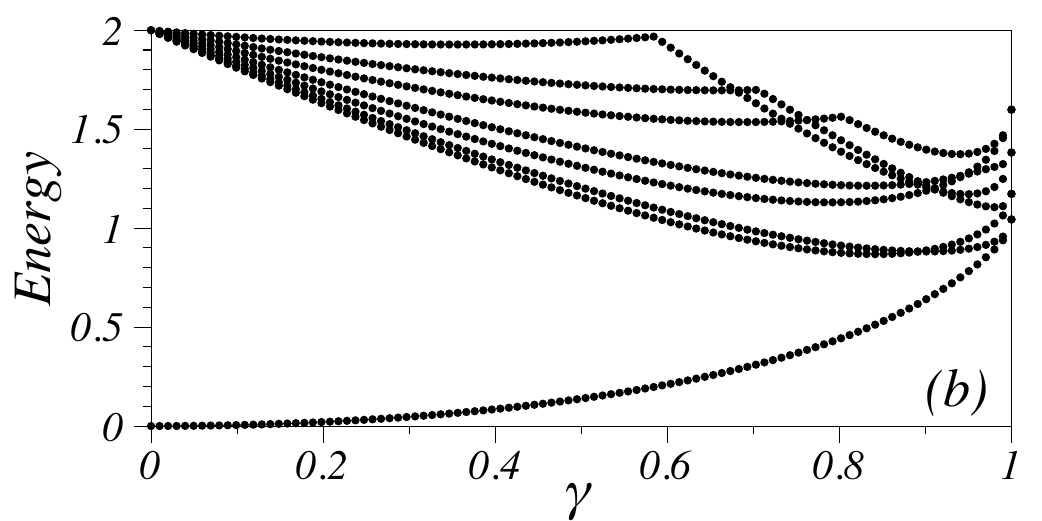}
\includegraphics[width=0.70\textwidth]{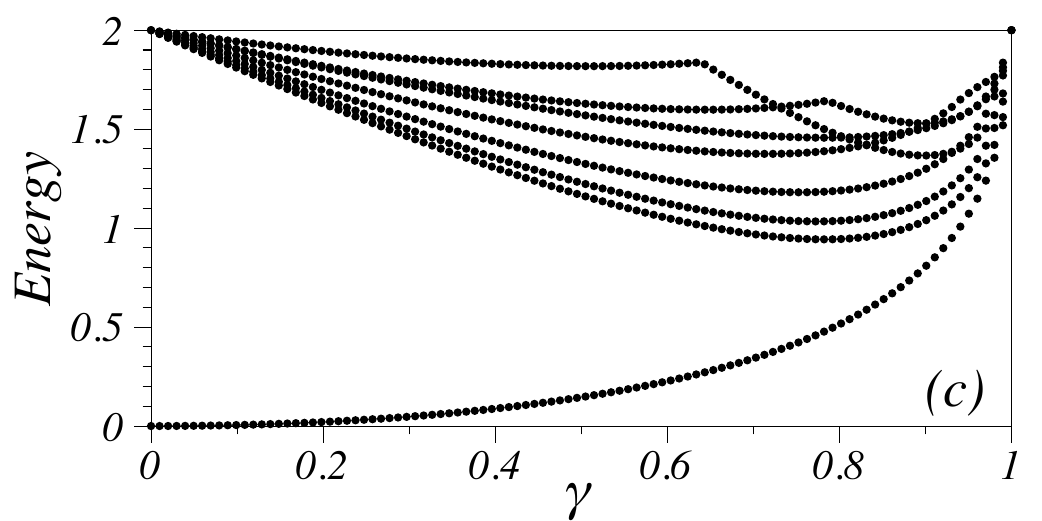}
\caption{Low energy states of the single flip Hamiltonians studied in Ref.\ \cite{Augusiak10}
as a function of the temperature parameter $\gamma=\tanh 2\beta J$ for the same
parameters as in figure \ref{fig:spectra}. The three panels correspond to the same $\tau$-spin configurations,
even though the variables $\tau$ are defined differently.
Notice that in this case the spectra becomes degenerate always at $\gamma=1$,
and that the ground state is always unique.
 } \label{fig:spectraOneSpin}
\end{center}
\end{figure}

 Let us shortly comment on the above model. First, it is easy to
 notice that for $\tau=0$ or $\tau=2^{N}-1$, from Eqs.\
 (\ref{singleHamil}), (\ref{Ai}), and (\ref{Bi}) one recovers the
 Hamiltonian (\ref{HamilFelderhof}). Since, as shown in Ref.
 \cite{Felderhof71} the Hamiltonian (\ref{HamilFelderhof}) has a
 ground state with zero energy, it means that there exist
 off-diagonal elements surviving the evolution. On the other hand
 for $\tau\neq 0,2^{N}-1$ one gets (\ref{HamilFelderhof}), however,
 with some impurities. After substitution of bond variables (see
 e.g. \cite{Siggia77}) one can map $H_{\tau}$ to disordered
 Heisenberg chains meaning that for some particular values of the
 involved parameters the model can be solved analytically. On the
 other hand, one may always treat this model numerically through
 matrix product states.

We show in Fig.\ \ref{fig:spectraOneSpin} the spectra for some
of these Hamiltonians, which is to be contrasted with the spectra
from the two spin flip models, Fig.\ \ref{fig:spectra}.
In the single flip model  the ground state is always unique except at zero temperature, where
for all of the associated Hamiltonians one observes criticality.

\section{Discussion and Outlook}

In these lectures we have seen how quantum information theory
can bring about a fresh perspective into many-body physics.
However, the field is much bigger than what we have reviewed.
Let us just mention here a few relevant topics that we have not covered,
but that have received plenty of attention from the community, and that
certainly have contributed to sizable advances in our understanding of many
body physics.

One interesting application of entanglement is to critical
phenomena. We briefly saw how block entanglement entropy scales
differently at a gapless critical point. However, many
entanglement measures display some kind of special behavior around
quantum criticality --- which was first observed \cite{Osterloh02}
in the concurrence of nearest-neighbor spins of an Ising chain
(see Ref.\ \cite{Amico08} for a recent review of activity in this
field). Quantum criticality, in fact, is a very active subject in
the condensed matter community, and has been studied using other
quantum information approaches like the ground state fidelity
\cite{Zanardi06} and the Loschmidt echo \cite{Quan06}, whose
usefulness in practice has been demonstrated experimentally
\cite{Zhang08,Zhang09}.

Another problem that is gaining interest is that of topological
order, which we mentioned briefly as one of the motivations for
studying area laws. One interesting recent development is the
study of entanglement spectra \cite{Haldane08,Calabrese08},
defined through the Schmidt decomposition in such a way that each
Schmidt coefficient $\lambda_\alpha$ of a bipartition is
interpreted as a dimensionless energy $\xi_\alpha=-\log
\lambda_\alpha$. This approach allows to generalize the von
Neumann block entropy by introducing a virtual temperature, and
study the structure of entanglement with more detail. In
particular, it appears that gapped systems with topological order
always have a gapless entanglement spectrum \cite{Haldane08}. As a
characterization of entanglement, the whole spectrum promises to
be better than just entropy --- simply because a set of numbers
contains more information than a single one.

Although we concentrated mostly on the theoretical aspect of
matrix and tensor product states, the field is also strongly
geared to the practical application of simulation of many-body
systems in classical computers. On the theory side, the tensor
product approach has given successful advances in the theory of
computational complexity applied to quantum mechanics
\cite{Schuch07}, and in the recent theory of entanglement
renormalization \cite{Vidal07}. 

On the computational side, MPS
algorithms have expanded the effective DMRG methods, and
tremendous progress is being done in the simulation of strongly correlated
particle systems. Bosonic particles can be represented straightforwardly\cite{Clark04} 
by mapping the $d$ internal levels of the spins 
to the occupation number at each lattice site -- thus truncating the Hilbert space
to the subspace with at most $d-1$ particles in each site.
Fermionic models, however, require some extra care
when contracting the indices in the network so that fermionic commutation relations
are respected \cite{Kraus10,Corboz09,Corboz10,Barthel09,Corboz10b,Pineda10}. 
In any case,  tensor networks are still computationally efficient with
 strongly correlated electron systems, which puts these algorithms at an advantage over quantum
 Montecarlo type techniques -- who suffer from the so called ``sign problem'' in this
 type of systems \cite{Troyer05}.
Therefore, tensor network techniques are important to support the
 large experimental efforts towards implementing quantum simulations of fermionic models
 (mainly with trapped ions \cite{Kim10} and ultracold atomic systems \cite{Tarruell10}).

The quantum kinetic Ising models discussed in the last section
hold plenty of potential for the near future in at least two
fronts. First, they represent a whole new class of many-body
systems amenable to analytical solution, and can therefore bring
new insight into our understanding of complex quantum many-body
dynamics, as well as some classical reaction-diffusion
problems \cite{Temme09}. Second, they have a close relationship and
could be useful to the recent ideas on ``environment design''
\cite{Verstraete09,Kraus08,Diehl08}: crafting and/or manipulating
the environment of a system so that it is driven to an interesting
quantum many-body state, usually with a dynamics given by a
quantum master equation. Because quantum kinetic models can be
well understood and controlled, they might provide the foundation
on top of which more elaborated systems are designed.

\begin{acknowledgement}
We are grateful to Ll.\ Masanes for helpful discussion.
We acknowledge the
support of Spanish MEC/MINCIN projects TOQATA (FIS2008-00784) and
QOIT (Consolider Ingenio 2010), ESF/MEC project FERMIX
(FIS2007-29996-E), EU Integrated Project  SCALA, EU STREP project
NAMEQUAM, ERC Advanced Grant QUAGATUA, Caixa Manresa, AQUTE, and
Alexander von Humboldt Foundation Senior Research Prize.
\end{acknowledgement}


\end{document}